\documentclass[
a4paper,															
11pt,		 														
parskip=on,  														
oneside, 															
onecolumn,															
numbers=noenddot,													
bibliography=totoc,													
listof=totoc,														
]{scrartcl} 														

\usepackage[german, english]{babel} 			

\usepackage[utf8]{inputenc}						
																				
\usepackage{babelbib}							

\usepackage[T1]{fontenc}		        		
												
\usepackage{lmodern}							

\usepackage{amsmath} 					  		

\usepackage{amssymb} 							

\usepackage[version=3]{mhchem}					

\usepackage{units} 								

\usepackage{float}								

\usepackage{graphicx}			          		

\usepackage[figuresright]{rotating}				

\usepackage{xspace} 							

\usepackage[dvipsnames]{xcolor}					

\usepackage[numbers, sort&compress]{natbib}		

\usepackage{multicol} 						

\usepackage[verbose]{placeins} 				

\usepackage[babel]{microtype} 				

\usepackage[								
left=28mm,  
right=25mm,	
bottom=25mm,								
top=28mm									
]{geometry} 

\usepackage{fancyhdr}
\usepackage{cite}

\usepackage[labelfont=bf]{caption} 	

\usepackage{setspace}
\usepackage{boldline}	 			

\usepackage{siunitx}
\usepackage{lscape} 		
\usepackage{pdflscape}		
\usepackage{tablefootnote} 	
\usepackage{multirow} 		
\usepackage{longtable}	
\usepackage{arydshln} 		
\newcolumntype{C}[1]{>{\centering\arraybackslash}p{#1}}
\usepackage{booktabs}		
\usepackage{bm} 			
\usepackage{mathtools} 		
\usepackage{supertabular}
\usepackage{tabularx} 		
\usepackage[para]{threeparttable}
\usepackage{etoolbox} 
\AtEndEnvironment{threeparttable}{\vskip10pt}{}{}

\allowdisplaybreaks 	

\usepackage{bookmark}	

\write18{} 	
\usepackage{tikz}
\usepackage{pgfplots} 	
\pgfplotsset{compat=newest} 
\usepgfplotslibrary{external,polar,patchplots}
\tikzset{external/force remake} 

\usetikzlibrary{
	arrows,
	patterns,
	calc,
	shapes.arrows,
	intersections,
	decorations,
	shapes,
	positioning,
	shadings,
	plotmarks,
	pgfplots.colormaps, 
	spy 		
} 

\newcommand{\cotwo}{CO\textsubscript{2}\xspace}		
\newcommand{\chfour}{CH\textsubscript{4}\xspace}	

\newcommand{\pd}{\partial}	

\usepackage{makecell} 		

\usepackage{hyphenat}
\begin{document}
																												

\pagestyle{fancyplain} 	
\fancyhf{} 				
\chead{ \fancyplain{}{This work has been accepted for the 24\textsuperscript{th} Intl. Conference on Process Control.
} } 	

\begin{center}
\huge Observability and Identifiability Analyses of Process Models for Agricultural Anaerobic Digestion Plants \\ [5mm]

\Large
Simon Hellmann\textsuperscript{1}, Arne-Jens Hempel\textsuperscript{2}, Stefan Streif\textsuperscript{3}, Sören Weinrich\textsuperscript{4,1} \\ [3mm]

\large
\textsuperscript{1} DBFZ Deutsches Biomasseforschungszentrum gGmbH, Leipzig, Germany,\\
\{simon.hellmann, soeren.weinrich\}@dbfz.de\\ [3mm]

\textsuperscript{2} Saxon University of Cooperative Education, Glauchau, Germany,\\ 
arne-jens.hempel@ba-sachsen.de \\ [3mm]

\textsuperscript{3} Chemnitz University of Technology, Chemnitz, Germany,\\ 
stefan.streif@etit.tu-chemnitz.de\\ [3mm]

\textsuperscript{4} Münster University of Applied Sciences, Faculty of Energy $\cdot$ Building Services $\cdot$ Environmental Engineering, Münster, Germany, weinrich@fh-muenster.de
\end{center}

\Large
This is the extended version of an accepted paper submitted to the 24th International Conference on Process Control on April 4, 2023. The extended version is available under a CC BY-NC-ND 4.0 license.

\vfill
\normalsize

\copyright 2023 IEEE. Personal use of this material is permitted. Permission from IEEE must be obtained for all other uses, in any current or future media, including reprinting/republishing this material for advertising or promotional purposes, creating new collective works, for resale or redistribution to servers or lists, or reuse of any copyrighted component of this work in other works.

\newpage
\pagestyle{plain} 		
\setcounter{page}{1}
\cfoot[]{\thepage}		

\section*{Abstract}
\addcontentsline{toc}{section}{Abstract}
Dynamic operation of anaerobic digestion plants requires advanced process monitoring and control. 
Different simplifications of the Anaerobic Digestion Model No. 1 (ADM1) have been proposed recently, which appear promising for model-based process automation and state estimation. 
As a fundamental requirement, observability and identifiability of these models are analyzed in this work, which was pursued through algebraic and geometric analysis. Manual algebraic assessment was successful for small models such as the ADM1-R4 and simplified versions of the ADM1-R3, which were derived in this context. However, for larger model classes the algebraic approach showed to be insufficient. By contrast, the geometric approach, implemented in the STRIKE\_GOLDD toolbox, allowed to show observability for more complex models (including ADM1-R4 and ADM1-R3), employing two independent algorithms. The present study lays the groundwork for state observer design, parameter estimation and advanced control resting upon ADM1-based models.

\begin{tabular}{rl}
	\textbf{Key words:} & ADM1, model simplification, Biogas technology, parameter estimation, \\ & STRIKE\_GOLDD
\end{tabular}

\clearpage

\section{Introduction}
%
Anaerobic digestion (AD) allows to convert numerous organic feedstocks into biogas, which can either be upgraded and injected into the natural gas grid or combusted to produce renewable electricity and heat. 

The AD process naturally shows strongly nonlinear behavior and is sensitive to process inhibition \citep{Chen2008}. Moreover, the process is prone to instability, especially during dynamic feeding \citep{Gaida2017}. To avoid instable process behavior, monitoring and control schemes are required.

Many investigations focus on automated operation of domestic or industrial wastewater treatment plants \citep{FloresEstrella.2019, AlcarazGonzalez2021, LaraCisneros2016, RochaCozatl.2015}. 
Additionally, advanced control for efficient and demand-oriented biogas production of agricultural AD plants are frequently examined \citep{Mauky.2016, Gaida2017, Dewasme.2019}. 

A bottleneck in full-scale application of AD is the lack of online measurements, especially regarding reliable stability indicators, such as volatile fatty acids (VFA) and alkalinity \citep{Kazemi2020}. A remedy to overcome this shortage is to apply soft sensors (or state observers), which use readily available external measurements and a mathematical model of the process to estimate internal, non-measurable process states \citep{Jimenez2015}. 

A prominent process model was presented by Bernard et al. \citep{Bernard2001}, which 
has successfully been used in multiple monitoring and control applications \citep{Attar.2018, FloresEstrella.2019, AlcarazGonzalez2021}. 

However, nonlinear aspects of the AD process are described in more detail within the established Anaerobic Digestion Model No. 1 (ADM1) 
\citep{Batstone2002}. 
While the model proposed by Bernard et al. only includes pH inhibition, the ADM1 covers process inhibition through pH, nitrogen limitation and ammonia. 
Still, successful applications of the ADM1 in full-scale monitoring and control applications have not yet been reported, mostly because of its complexity and vast number of parameters \citep{DonosoBravo.2011}. 

Yet, in an agricultural setting, which is an important application of AD, both the orginal ADM1 and the model by Bernard et al. cannot directly be used due to their underlying reference unit (chemical oxygen demand, COD) \citep{Weinrich2021b}. Thus, Weinrich and Nelles have recently proposed mass-based simplifications of the ADM1 \citep{Weinrich2021b, Weinrich2021}. These models represent a suitable alternative for application in monitoring and control of agricultural AD plants \citep{Mauky.2016}. 
Individual model variations differ significantly in their number of differential equations, state variables and required parameters (Fig.~\ref{fig:meth:modelSimplificationsSoeren}). Simplified models (such as the ADM1-R4) combine nutrient degradation and biogas formation based on first-order sum reactions, whereas more detailed models (such as the ADM1-R3) depict specific degradation pathways and ammonia or pH inhibition during acetoclastic methanogensis.

%
Observability is a model property which indicates whether internal states can 
be inferred based on input-output measurements \citep{Anguelova.2007}. Likewise, identifiability implies that model parameters can be calibrated based on input-output measurements. Assessing observability and identifiability is therefore a fundamental requirement before implementing state and parameter estimation. 

Numerous approaches for assessing observability and identifiability have been proposed \citep{Villaverde.2016, Maes.2019, Chis.2011}. However, especially for complex models they are seldom analyzed a priori because of the computational complexity of the symbolic calculations involved \citep{Chis.2011}. 

This contribution analyzes observability and identifiability of different ADM1 simplifications proposed by Weinrich and Nelles \citep{Weinrich2021b}. For this purpose, two different approaches are pursued: the differential algebraic and differential geometric approach.\footnote{The two approaches are simply referred to as algebraic and geometric in the following.} Typical measurements at full-scale AD plants are assumed to be available to ensure feasibility of future process control schemes based on these analyses.
%
%
%
%

\section{Methods}
In this work, we consider systems of ordinary differential equations of the form:
%
%
\begin{equation}
	M : \left\{
		\begin{aligned}
			\dot{x} \quad &= f\left(x(t),u(t),\theta\right) \\
			y \quad &= h\left(x(t),\theta\right) \\
			x(t_0) &= x_0
		\end{aligned}
	\right.
\end{equation}
with state variables $x \in \mathbb{R}^{n}$, initial state $x_0$, manipulated variables $u \in \mathbb{R}^{p}$, measurement variables $y \in \mathbb{R}^{q}$, and model parameters $\theta \in \mathbb{R}^{m}$. Generally, model parameters (such as microbial growth or decay rates) can be 
time-variant. However, they vary at a slow rate of change and their dynamics do not follow a defined differential equation. Therefore, their implicit time dependence is suppressed in the notation.
\subsection{Modelling of the Anaerobic Digestion Process}
\label{sec:meth:ADModels}
%
%
\begin{figure}[b!t]
	\centering
	\includegraphics[width=\linewidth]{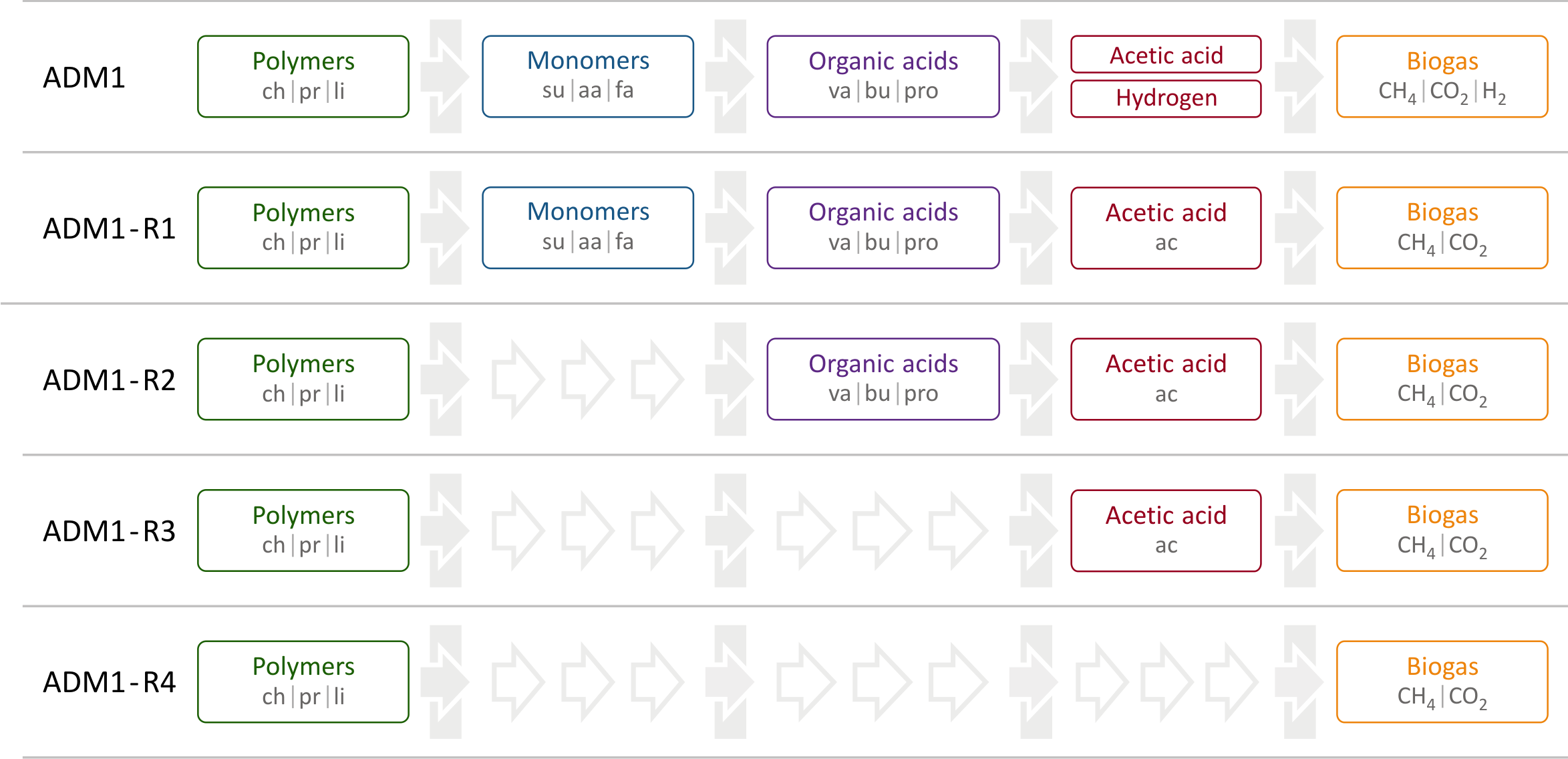}
	\caption{Characteristics of different ADM1 simplifications \citep{Weinrich2021b}.}
	\label{fig:meth:modelSimplificationsSoeren}
\end{figure}
%
Based on available model simplifications in Fig.~\ref{fig:meth:modelSimplificationsSoeren}, observability and identifiability were assessed for the ADM1-R4, ADM1-R3 and ADM1-R2 \citep{Weinrich2021b}. 
Mass concentration of ash was integrated to compute volatile solids (VS) measurements. 
\subsubsection{Model Simplification}
\begin{figure}[b!]
	\centering
	\includegraphics[width=0.95\linewidth]{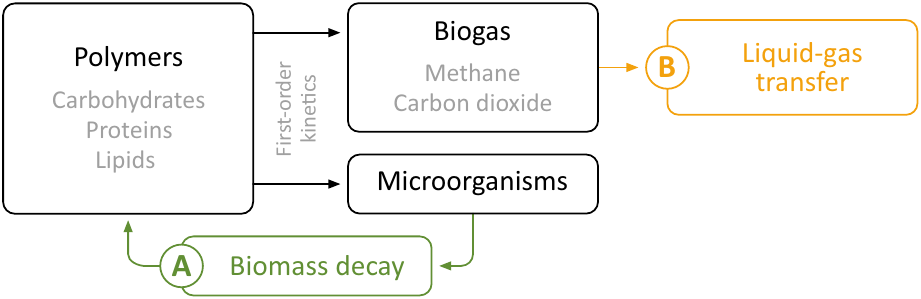}
	\caption{Model components of the ADM1-R4.}
	\label{fig:meth:modelVariantsR4}
\end{figure}

Both ADM1-R4 and ADM1-R3 were further simplified throughout the investigation. For this purpose, individual model components were isolated and omitted or incorporated systematically to assess their influence on observability. 
Fig.~\ref{fig:meth:modelVariantsR4} illustrates the full ADM1-R4 and its model parts qualitatively.
These model parts are decay of microbial biomass and its stoichiometric feedback as macro nutrients (part~A, in green), and gas solubility of methane and carbon dioxide (part~B, in orange). 

\begin{figure}[b!]
	\centering
	\includegraphics[width=\linewidth]{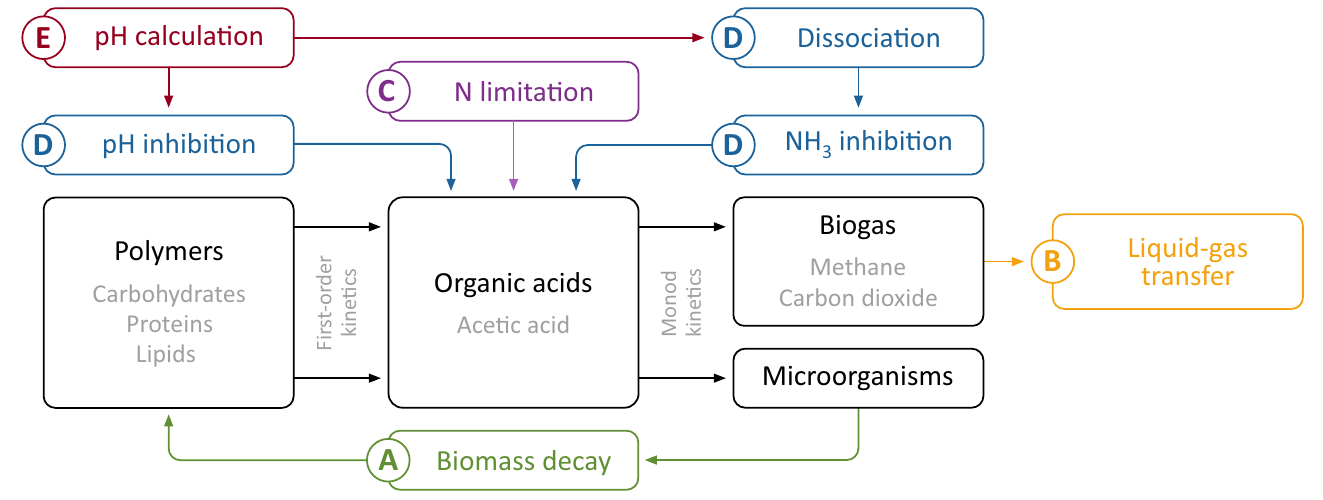}
	\caption{Model components of the ADM1-R3.}
	\label{fig:meth:modelVariantsR3}
\end{figure}

The ADM1-R3 allows to isolate more model parts as shown in Fig.~\ref{fig:meth:modelVariantsR3}. Part~A and B were left identical as for the ADM1-R4 (in green and orange, respectively). Part~C (in purple) describes inhibition through nitrogen limitation. Part~D (in blue) covers inhibition through pH and ammonia, as well as dissociation of ammonium/ammonia. Lastly, part~E (in red) contains the computation of pH, which includes the charge balance of available anions and cations. 

The core elements of the models ADM1-R4 and ADM1-R3 without any additional model parts are denoted as BMR4 and BMR3 (base~model, BM). Augmenting them with additional model parts results in e.g. BMR4+A. The same notation applies for individual ADM1-R3 model variations. The investigated models are summarized in Tab.~\ref{tab:res:algApproach}. A full set of the corresponding model equations is given in the appendix.
\subsubsection{Data Availability}
\label{sec:meth:dataAvailability}
For observability and identifiability analyses, measurement signals in Tab.~\ref{tab:meth:measurements} were considered. 
\begin{table}[b]
	\renewcommand{\arraystretch}{1.1}
	\caption{Measurement signals considered for individual model classes}
	\begin{center}
		\begin{tabular}{|l|l|l|}
			\hline
			\textbf{Model class} & \textbf{Online} & \textbf{Offline$^{\mathrm{a}}$} \\
			\hline
			ADM1-R4 & gas composition (CH\textsubscript{4}, CO\textsubscript{2}) & TS, VS, IN \\
			\hline 
			ADM1-R3$^{\mathrm{b}}$ & gas composition (CH\textsubscript{4}, CO\textsubscript{2}), pH & TS, VS, IN \\
			\hline
			\multicolumn{3}{l}{$^{\mathrm{a}}$Total solids (TS), volatile solids (VS) and inorganic nitrogen (IN)} \\
			\multicolumn{3}{l}{$^{\mathrm{b}}$During algebraic oberservability analyses acetic acid was also}\\
			\multicolumn{3}{l}{considered as an optional online measurement.}
		\end{tabular}
		\label{tab:meth:measurements}
	\end{center}
\end{table}
Online measurements were generally restricted to pH and gas composition, expressed as partial pressures of methane and carbon dioxide \citep{Wolf2014}. For detailed assessment of observability using the algebraic approach, acetic acid was also considered as available online. For lab and pilot-scale settings, this is a reasonable assumption \citep{Boe.2012}. Generally, however, online measurements of acetic acid were not considered, which represents the more realistic scenario in a full-scale agricultural setting \citep{Wolf2014}. 

Offline measurements of total solids (TS), volatile solids (VS) and inorganic nitrogen (\mbox{IN}) were assumed to be slowly time-variant in between samples \citep{Liebetrau2020}. To obtain time-continuous signals, sample-and-hold behavior was supposed.
%
\subsection{Observability Analyses}
\textbf{Definition.} \textit{A state variable $x_i(t)$ is observable if its initial state $x_i(0)$ can be reconstructed from measurements of inputs $u(\tau)$ and outputs $y(\tau)$ over a finite time $\tau \in [0,t]$. A system $M$ is fully observable if all of its $n$ states are observable. If at least one state is not observable, the system is not fully observable \citep{Simon.2006}.}


In this work, two pathways to assess observability were pursued: the algebraic and the geometric approach. In the former, a system of equations of output variables and their time derivatives is established and solved for individual model states. The latter assesses the observability rank condition, which relies on an observability matrix. 
\subsubsection{Algebraic Approach}
In the algebraic approach, a system of equations of output variables $y$ and their time derivatives $y, \dot y, \ddot y, \ldots$ is established and solved for the state variables $x_i, i \in 1, \ldots , n$. The system of equations can be summarized as
\begin{equation}
\label{eq:meth:equationSystemVector}
\mathcal{Y} = \mathcal{Y}(x,u,\theta)
\text{, where } 
\mathcal{Y} = \left(y, \dot y, \ddot y, \hdots\right)^{T}. 	
\end{equation}
Time derivatives of the outputs are obtained by iteratively computing the Lie derivatives 
with respect to $f$. We assumed constant input signals during individual feeding events ($\dot u = 0$) and constant parameters ($\dot \theta = 0$), resulting in:
%
%
\begin{eqnarray}
	y &=& h(x,\theta) \\
	\dot y &=& L_f h(x,\theta) = \frac{\pd h(x,\theta)}{\pd x} f(x,u,\theta) , \\
	\ddot{y} &=& L_f^2 h(x,\theta) = \frac{\pd L_f h(x,\theta)}{\pd x} f(x,u,\theta) , \\
	&\vdots&  \notag \\
	y^{(k)} &=& L_f^{(k)} h(x,\theta) = \frac{\pd L_f^{(k-1)} h(x,\theta)}{\pd x} f(x,u,\theta).
\end{eqnarray}
%

Observability is given if the system of equations \eqref{eq:meth:equationSystemVector} can be solved algebraically for all states $x_i, i=1\ldots n$ and has at least one solution \citep{Anguelova.2007}. This requires $n$ equations, which can either be obtained by incorporating all of the $q$ entries of each vector $y$, $\dot y$ etc.; or by building higher-order time derivatives $y^{(k)}$ and incorporating their elements. If the system of equations can be solved uniquely, global observability is shown. Multiple solutions indicate local observability \citep{Anguelova.2007}.
\paragraph*{Implementation}
The algebraic approach was implemented in Mathematica (version 13.0, Wolfram Research, Inc.). Equations which could be solved for state variables manually were excluded from the system of equations to minimize computational demand. Complexity of the system of equations was further reduced by seeking to incorporate terms with minimal orders of time derivatives first. This is because generally complexity of symbolic derivatives $y^{(k)}$ increases rapidly with the order of derivatives $k$ \citep{Chis.2011}. Selection of terms among the $q$ elements of $y^{(k)}$ followed a heuristic procedure.

Since the algebraic approach turned out to be inconclusive for the ADM1-R3, the models ADM1-R4 and ADM1-R3 were systematically simplified (see Sec.~\ref{sec:meth:ADModels}). The maximum allowable model complexity for the algebraic approach was determined by systematically adding model components, starting from the base models (BMR4 and BMR3), and terminating when the system of equations could no longer be solved. 

In the algebraic approach, all parameters and influent concentrations were assumed to be known and time-invariant. Moreover, for the ADM1-R3 model classes, measurements of acetic acid were assumed to be available online in order to achieve conclusive statements on observability, see Sec. \ref{sec:res:algApproach}.
\subsubsection{Geometric Approach}
In the differential geometric approach observability is investigated by computing the rank of an observability matrix. Two different algorithms were considered in this context, which differ in the way the observability matrix is built: Full Input-State-Parameter Observabilty 
(FISPO) and Observability Rank Condition with Direct Feedthrough (\mbox{ORC-DF}). Both algorithms rely on Lie derivatives of the output. For better understanding, computation of the observability matrix is explained by means of the FISPO algorithm. In a second step, disparities with respect to the \mbox{ORC-DF} algorithm are illustrated. 

In FISPO, the observability matrix $\mathcal{O}(x)$ is built by taking the Lie derivatives of the output symbolically and computing their partial derivatives with respect to the states $x$:
\begin{equation}
\label{eq:meth:fispoMatO}
\mathcal{O} ( x) = 
\left(\begin{array}{c} 
\frac{\pd h( x)}{\pd x} \\ \frac{\pd}{\pd x} \left(L_f h( x)\right) \\ \vdots \\ \frac{\pd}{\pd x} \left(L_f^{n-1} h( x)\right) 
\end{array}\right). 
\end{equation}
A model is locally observable 
in a neighborhood $N( x_0)$ of a point $ x_0$ if it holds that \citep{Villaverde.2016}
\begin{equation}
\text{rank} \left(\mathcal{O}( x_0)\right) = n.
\end{equation}
The ORC-DF algorithm pursues a different approach in computing the observability matrix and is only applicable for input-affine systems \citep{Maes.2019}, which can be described as:
%
\begin{eqnarray}
	\dot x &=& f(x) + \sum_{j=1}^{p} g_j (x) u_j, \label{eq:meth:ORCStateOde}\\
	y &=& h_0 (x) + \sum_{j=1}^{p} h_j(x) u_j.
\end{eqnarray}
$f, g_j \in \mathbb{R}^n$ and $h_0, h_j \in \mathbb{R}^p$ are vectorial functions, which describe the output $y$ uniquely. They are aggregated in a column vector $\Omega$. The observability matrix in ORC-DF is built repeatedly using symbolic computation. 
Initially, it is computed as partial derivatives of $\Omega$ with respect to the states $x$. If the observability rank condition can be satisfied, the system is locally observable. If not, the vector $\Omega$ is extended by the Lie derivatives of the previous version of $\Omega$ with respect to $f$ and $g_j$. This procedure continues until the rank condition is satisfied or a maximum number of iterations is reached. Details can be found in \citep{Maes.2019}. 

\paragraph*{Implementation}
Both geometric approaches were implemented in the Matlab toolbox STRIKE\_GOLDD 
4.0 \citep{DiazSeoane.2022}. The toolbox 
allows to assess local observability (and structural local identifiability) 
with the FISPO and ORC-DF algorithms.  

To ensure that statements on observability of the models are independent from the numeric value of the initial state~$x_0$, no previously known initial states $x_0$ were assumed for both procedures. This results in symbolic computations of $\mathcal{O}(x)$ and its rank.

In the geometric approach, full model classes ADM1-R4 and ADM1-R3 as well as all submodels were analyzed. 
Inlet concentrations were assumed to be known and time-invariant. 

Computations of both algebraic and geometric approaches were conducted on a standard personal computer.\footnote{Intel Core i5 processor (\SI{1.7}{GHz}), \SI{32}{GB} RAM and Windows 10 operating system.}
\subsection{Identifiability Analyses}
\label{sec:identifiability}
\textbf{Definition.} \textit{A model parameter $\theta_i, i=1,\ldots,m$ is structurally identifiable (s.i.) if for almost all true parameter values~$\theta^*$ the parameter estimate $\hat \theta$ can be determined from input-output behavior of the model $M$ \citep{Walter.1997}:}
\begin{subequations}
	\label{eq:meth:definitionIdentifiability}
	\begin{align}
	M\big( x(t), u(t), \hat \theta \big) &= M\big( x(t), u(t), \theta^* \big) \\
	\Rightarrow \hat \theta &= \theta^*.
	\end{align}
\end{subequations}
\textit{If all model parameters $\theta_i$ are s.i., the model is s.i.. A parameter $\theta_i$ is locally structurally identifiable (l.s.i.) if \eqref{eq:meth:definitionIdentifiability} holds in a neighborhood $\mathcal{N}(\theta^*)$. 
A full model is l.s.i. if all its parameters $\theta_i$ are l.s.i.. If at least one of them is not, the model is not l.s.i..}


Identifiability was analyzed as part of the differential geometric approach and considered as augmented observability \citep{Villaverde.2016, Maes.2019}. For this purpose, the state vector is augmented by the model parameters, for which trivial dynamics are assumed:
\begin{equation}
\label{eq:meth:stateAugmentation}
\dot {\tilde x} = 
\left(\begin{array}{c} 
\dot x \\
\dot \theta 
\end{array}\right) = 
\left(\begin{array}{c} 
f(x,u,\theta) \\
0 
\end{array}\right).
\end{equation}
Consequently, for FISPO, computation of the extended observability matrix is augmented to:
\begin{equation}
\label{eq:meth:fispoMatO_ext}
\mathcal{O} (\tilde x) = 
\left(\begin{array}{c} 
\frac{\pd h(\tilde x)}{\pd \tilde x} \\ \frac{\pd}{\pd \tilde x} \left(L_f h(\tilde x)\right) \\ \vdots \\ \frac{\pd}{\pd \tilde x} \left(L_f^{n+m-1} h(\tilde x)\right) 
\end{array}\right). 
\end{equation}
The system is considered observable and identifiable if the rank conditions satisfies \citep{Villaverde.2016}
\begin{equation}
\text{rank} \left(\mathcal{O}(\tilde x_0)\right) = n + m.
\end{equation}
For ORC-DF the state differential equations \eqref{eq:meth:ORCStateOde} are augmented by a zero vector of dimension $m \times 1$, accounting for the trivial dynamics of the parameters. The output equations remain unchanged. 

For the ADM1-R4, the rate constants of hydrolysis and decay were assumed as time-variant and analyzed for identifiability. For the ADM1-R3, the following parameters were investigated: maximum growth rate and half-saturation constant of acetoclastic methanogens as well as the inhibition constant for non-competitive ammonia inhibition.
%
%

\section{Results and Discussion}
This section presents the results of both the algebraic and geometric approach. Results of the full model variants are discussed first, followed by the simplified submodels. 
\subsection{Algebraic Approach}
\label{sec:res:algApproach}
The full ADM1-R4 (BMR4+AB) could be shown to be globally observable. Tab.~\ref{tab:res:algApproach} summarizes the required measurements, time derivatives and computing times for all succesful runs of the algebraic approach. Color codes of the submodels are consistent with Fig.~\ref{fig:meth:modelVariantsR4} and \ref{fig:meth:modelVariantsR3}. Note that alternative combinations to form the equation systems are possible and result in slightly different computing times. This applies for all successful runs of the algebraic approach. For the ADM1-R4, the underlying system of equations involves online measurements of the gas composition (CH\textsubscript{4} and CO\textsubscript{2}). 
Furthermore, to guarantee observability, measurements of IN, TS and VS need to be available. This follows directly from the model equations: all three states only appear in their corresponding differential equation. 
Therefore, if they were not available as measurements, they could not be observable because they would not be introduced into the system of equations via other measurements, regardless of the degree of time derivatives. 
\begin{table}[t]
	\scriptsize
	\vspace*{-7pt}
	\renewcommand{\arraystretch}{1.5}
	\caption{Measurements and their time derivatives required for algebraic observability of model simplifications}
	\label{tab:res:algApproach}
	\centering
	\setlength\tabcolsep{3pt}
	\begin{tabular}{|l|r|r V{3} *{5}{c|}c V{3} *{3}{c|}c V{3} *{3}{c|}c V{3} c|c|}
		\hline
		\multicolumn{3}{|l V{3}}{} & \multicolumn{6}{c V{3}}{\textbf{Nominal measurements}$^{\mathrm{a,b}}$} & \multicolumn{4}{c V{3}}{\textbf{1\textsuperscript{st} derivative}} & \multicolumn{4}{c V{3}}{\textbf{2\textsuperscript{nd} derivative}} & \multicolumn{2}{c|}{\textbf{3\textsuperscript{rd} deriv.$^{\mathrm{e}}$}} \\
		\hline
		\textbf{Model name} & $n^{\mathrm{c}}$ & $t [s]^{\mathrm{d}}$ & CH\textsubscript{4} & CO\textsubscript{2} & IN & TS & VS & Ac & CH\textsubscript{4} & CO\textsubscript{2} & IN & TS & CH\textsubscript{4} & CO\textsubscript{2} & IN & TS & CH\textsubscript{4} & CO\textsubscript{2} \\
		\hline
		BMR4 	& 9 & 5 	& x & x & x & x & x & 	& x & x & & & x & x & & & & \\
		\hline
		BMR4+\color{ForestGreen}{A} 	& 9 & 3 	& x & x & x & x & x & 	& x & x & & & x & x & & & & \\
		\hline
		BMR4+\color{Dandelion}{B} 	&11 & 6 	& x & x & x & x & x & 	& x & x & & & x & x & & & x & x \\
		\hline
		BMR4+\color{ForestGreen}{A}\color{Dandelion}{B} &11 & 8 	& x & x & x & x & x & 	& x & x & & & x & x & & & x & x \\
		\specialrule{.12em}{0.01em}{0.01em} 
		BMR3$^{\mathrm{f}}$ & 11 & 541 & x & x & x & x & x & x & x & x & x & x & x & & & & & \\
		\hline
		BMR3+\color{ForestGreen}{A}\color{black}$^{\mathrm{f}}$ & 11 & 291 & x & x & x & x & x & x & x & & x & x & x & & & x & & \\
		\hline
		BMR3+\color{ForestGreen}{A}\color{Orchid}{C}\color{black}$^{\mathrm{f}}$ & 11 & 1065 & x & x & x & x & x & x & x & & x & x & x & & & x & & \\
		\hline
		BMR3+\color{Dandelion}{B}\color{Orchid}{C}\color{black}$^{\mathrm{f}}$ & 13 & 5192 & x & x & x & x & x & x & x & x & x & x & x & x & & x & & \\
		\hline
		BMR3+\color{ForestGreen}{A}\color{Dandelion}{B}\color{Orchid}{C}\color{black}$^{\mathrm{f}}$ & 13 & 10225 & x & x & x & x & x & x & x & x	& & x & x & x & & & x & x \\
		\hline
		\multicolumn{19}{l}{$^{\mathrm{a}}$CH\textsubscript{4}, CO\textsubscript{2} - partial pressures of methane and carbon dioxide; IN - inorganic nitrogen; TS, VS - total and volatile} \\
		\multicolumn{19}{l}{solids; Ac - acetic acid \quad $^{\mathrm{b}}$CH\textsubscript{4} and CO\textsubscript{2} were assumed as online measurements, as well as Ac for ADM1-R3 } \\
		\multicolumn{19}{l}{variants. IN, TS and VS were assumed as offline measurements. \quad $^{\mathrm{c}}$Number of states \quad $^{\mathrm{d}}$Computation time}\\
		\multicolumn{19}{l}{$^{\mathrm{e}}$derivative \quad $^{\mathrm{f}}$Only local observability could be shown because two solutions to the equation systems were found.}
	\end{tabular}
\end{table}

Model simplifications were applied to the ADM1-R4, resulting in the submodels BMR4, BMR4+A and BMR4+B. This was done to explore the effect of individual model parts on the complexity of the equation system, indicated by the computing time required to solve the system of equations.   
Neglecting model part B reduced the computing time the most (BMR4+A). Only neglecting model part A (BMR4+B) also reduced computing time, albeit not as much. However, neglecting both model parts A and B (BMR4) resulted in slightly longer computing times than for BMR4+A. This is reasonable because simplifying the stoichiometric degradation pathway for biomass decay (model part A) in fact results in more complex terms for the time derivatives of the gas partial pressures if gas solubility (model part B) is neglected as well.  

For the full ADM1-R3 (equivalent to BMR3+ABCDE), the solver of Mathematica failed to deliver conclusive statements on observability, likely due to the complexity of the resulting 
terms in the system of equations. Neither a set of solutions nor an empty set could be returned. Instead, the kernel systematically died in the process of solving the equation systems. Hence, with the algebraic approach, observability of the full ADM1-R3 could not be determined. Therefore, model complexity of the ADM1-R3 was reduced by isolating and omitting individual model parts (see Sec.~\ref{sec:meth:ADModels}). 

Up to this point, measurements of CH\textsubscript{4}, CO\textsubscript{2}, IN, TS and VS were assumed. Additionally, for all ADM1-R3 submodels online measurements of acetic acid had to be taken into account to show local observability, see Tab.~\ref{tab:res:algApproach}. The obtained results are therefore strictly valid only for lab- and pilot-scale settings \citep{Boe.2012}. 

With given nominal measurements of VS, it was preferred to include time derivatives of TS over VS because both terms added the same information to the system of equations, but TS delivered less complex terms. 

BMR3+ABC is the most complex model variant whose resulting system of equations could still be solved with the algebraic approach, see Tab.~\ref{tab:res:algApproach}. Two solutions of the systems of equations were obtained, indicating local observability. 

For both BMR3+AC and BMR3+BC, local observability could be shown because two solutions each were obtained. 
Neglecting model part B reduced the computing time more than neglecting model part A, which is in line with the \mbox{ADM1-R4} submodels.

When further reducing BMR3+AC by model part C (BMR3+A), the shortest computing times of all ADM1-R3 submodels was achieved. However, omitting all model parts (BMR3) resulted in increased computing time compared with BMR3+A. This is consistent with the findings from the ADM1-R4 submodels.


Higher-order variants than BMR3+ABC result in an increased number of states because computation of pH needs to be considered as well as inhibition through pH and ammonia. This leads to symbolic systems of equations too complex to be solved in the presented framework. 

When restricting measurements 
to gas composition, IN, TS and VS (no measurement of acetic acid), none of the ADM1-R3 submodels could be shown to be observable with the algebraic approach because complexity of the systems of equations increased significantly. This is comprehensible since in this case the state variable of acetic acid needs to be incorporated through higher-order time derivatives of the other measurement equations.

%
\subsection{Geometric Approach}
Based on the implementation in the STRIKE\_GOLDD toolbox, the geometric approach proved to be successful in showing both local observability and structural local identifiability of the full models ADM1-R4 and ADM1-R3 and all of their submodels. Moreover, this was achieved in (i) shorter computing times, (ii)~without assuming acetic acid as an online measurement, and (iii) by employing two independent algorithms (FISPO and ORC-DF). Tab.~\ref{tab:res:geomApproach} summarizes the computing times for the full models ADM1-R4 and  \mbox{ADM1-R3} as well as for BMR3-ABC, being the most complex model that could successfully shown to be observable with the algebraic approach, cf. Tab.~\ref{tab:res:algApproach}. 
As noted by \citep{Martinez.2020}, computational efficiency of FISPO and ORC-DF can differ vastly depending on the model structure, which is especially apparent for the full ADM1-R3. 

Furthermore, computational effort decreased significantly when neglecting model parts of the full ADM1-R3: computing times of the submodel BMR3-ABC reduced to 12 and \SI{5}{\second} for FISPO and ORC-DF, compared with about 12,000 and \SI{800}{\second} for the full ADM1-R3, respectively. 

Computation times of the geometric approach were generally shorter than for the algebraic one. This is reasonable because in the former only the existence of a solution to the systems of equations is assessed (through the obserbility rank condition), whereas the equation systems have to be solved expliciteply in the latter.

%
\begin{table}[tb]
	\renewcommand{\arraystretch}{1.1}
	\caption{Model properties and computing times of FISPO and ORC-DF algorithms for ADM1-R4, BMR3+ABC and ADM1-R3.}
	\centering
	\begin{tabular}{|l|r|r|r|r|}
		\hline
		\multirow{2}{*}{\textbf{Model class}} & \multicolumn{2}{c|}{\textbf{number of}} & \multicolumn{2}{c|}{\textbf{computing time in s}} \\
		\cline{2-5} 
	 & \textbf{states} & \textbf{parameters} & \textbf{FISPO} & \textbf{ORC-DF} \\
		\hline
		ADM1-R4 & 11 & 4 & 3 & 7 \\
		\hline 
		BMR3+\color{ForestGreen}{A}\color{Dandelion}{B}\color{Orchid}{C}& 13 & 7 & 12 & 5 \\
		\hline 
		ADM1-R3 & 17 & 7 & 11959 & 811 \\
		\hline
	\end{tabular}
	\label{tab:res:geomApproach}
\end{table}

Higher-order model classes such as the ADM1-R2 are of similar structure as the ADM1-R3, but involve significantly more time-variant parameters and contain a more detailled acid spectrum (acetic to valeric acid instead of only acetic acid) \citep{Weinrich2021b}, and thus more states. In an agricultural setting, these acid measurements are not available online. Even when assuming them to be available online, both algorithms of the geometric approach failed to evaluate the respective observability rank condition, and thus did not allow to draw conclusive statements. This is likely due to the aggravated complexity of the involved symbolic expressions. Similar behavior is therefore expected for even more complex model classes (ADM1-R1 or ADM1). In such cases, more advanced methods for assessing observability are required.

A practical application of model classes higher than the ADM1-R3 (e.g. ADM1-R2) is thus not anticipated for monitoring and control schemes due to the lack of online measurements of individual VFAs. It was therefore not further simplified into submodels and no further observability and identifiability analyses were pursued. 

\section{Conclusion}
The models ADM1-R4 and ADM1-R3 were analyzed for observability and identifiability using differential algebraic and geometric approaches.  
With the former, observability of the ADM1-R4 was successfully shown. However, the algebraic approach failed for the full ADM1-R3 if only typical measurements at agricultural biogas plants were considered. In this scenario though, the geometric approach succeeded to show observability and identifiability for both ADM1-R3 and ADM1-R4, and exhibited a higher computational efficiency. This emphasizes that the ADM1 simplifications of Weinrich and Nelles \citep{Weinrich2021b} are indeed promising for state and parameter estimation as well as process automation for agricultural AD plants.

\section*{Acknowledgment}
The authors are thankful for funding from German Federal Ministry of Food and Agriculture of the junior research group on simulation, monitoring and control of anaerobic digestion plants (grant 2219NR333). S. H. thanks Terrance Wilms for his encouragement and advice.

\appendix
\section*{Appendix}
\addcontentsline{toc}{section}{Appendix}
The ADM1 simplifications of Weinrich and Nelles \citep{Weinrich2021b} have been transformed into standard control notation. This section summarizes the model equations and parameters of all successfully investigated models, starting with the simplest model structures. The following notation was applied: 
\begin{itemize} \label{list:notation}
	\addtolength\itemsep{-5mm}
	\item $x_i$ - states (mass concentrations of the species involved), 
	\item $y_i$ - measurements
	\item $u$ - control variable (feed volume flow)
	\item $\xi_i$ - time-variant, uncertain parameters (inlet mass concentration)
	\item $\theta_i$ - time-variant parameters
	\item $c_i$ - (aggregated) time-invariant parameters 
	\item $a_{ij}$ - time-invariant stoichiometric coefficients
\end{itemize}
States and inlet concentrations are usually given in $\si{\gram\per\litre}$.\footnote{For improved numerical stability, it is advised to express the water concentration $S_\mathrm{h20}$ in \si{\kilogram\per\litre}. This levels out the large differences in numerical values present for a typically watery fermenter content (high values of $S_\mathrm{h20}$ compared with all other species).} The feed volume flow is stated in \si{\litre\per\day}. Units of states, inputs and measurements are summarized in Table~\ref{tab:app:units}. Units of all model parameters can be reviewed in \citep{SorenWeinrich2017}.
\begin{table}[htb]
	\renewcommand{\arraystretch}{1.1}
	\caption{Units of states, inputs and measurements of all described models.}
	\centering
	\begin{tabular}{cll}
		\toprule
		Symbol & Description & Unit \\
		\midrule
		$x_i$ & states & \si{\gram\per\litre} \\
		$\xi_i$ & inlet concentrations & \si{\gram\per\litre} \\
		$u$ & feed volume flow & \si{\litre\per\day} \\
		\midrule
		$\dot{V}_g$ & biogas volume flow & \si{\litre\per\day}\\ 
		$p_\mathrm{ch4}$ & partial pressure of methane & \si{\bar}\\ 
		$p_\mathrm{co2}$ & partial pressure of carbon dioxide & \si{\bar}\\
		$pH$ & pH value & $-$\\
		$S_\mathrm{IN}$ & inorganic nitrogen concentration & \si{\gram\per\litre}\\
		$TS$ & total solids content & $-$\\
		$VS$ & volatile solids content & $-$\\
		$S_\mathrm{ac}$ & acetic acid concentration & \si{\gram\per\litre}\\
		\bottomrule
	\end{tabular}
	\label{tab:app:units}
\end{table}

\subsection*{ADM1-R4 Models}
Aside from the ADM1-R4, the following models were analyzed: BMR4+B, BMR4+A, and BMR4. 
\subsubsection*{ADM1-R4}
State vector: 
\begin{equation}
\label{eq:R4:stateVector}
x= \left[S_\mathrm{ch4}, S_\mathrm{IC}, S_\mathrm{IN}, S_\mathrm{h2o}, X_\mathrm{ch}, X_\mathrm{pr}, X_\mathrm{li}, X_\mathrm{bac}, X_\mathrm{ash}, S_\mathrm{ch4,gas}, S_\mathrm{co2,gas}\right]^T
\end{equation}
Differential equations: 
\begin{subequations} \label{R4-controlNotation}
\begin{align}
\dot x_1 &= c_1 \left(\xi_1 - x_1\right) u + a_{11} \theta_1 x_5 + a_{12} \theta_2 x_6 + a_{13} \theta_3 x_7 - c_2 x_1 + c_3 x_{10} \label{R4-x1-control}\\
\dot x_2 &= c_1 \left(\xi_2 - x_2\right) u + a_{21} \theta_1 x_5 + a_{22} \theta_2 x_6 + a_{23} \theta_3 x_7 - c_2 x_2 + c_4 x_{11} \label{R4-control-x2}\\
\dot x_3 &= c_1 \left(\xi_3 - x_3\right) u - a_{31} \theta_1 x_5 + a_{32} \theta_2 x_6 - a_{33} \theta_3 x_7 \\
\dot x_4 &= c_1 \left(\xi_4 - x_4\right) u - a_{41} \theta_1 x_5 - a_{42} \theta_2 x_6 - a_{43} \theta_3 x_7 \\ 
\dot x_5 &= c_1 \left(\xi_5 - x_5\right) u - \theta_1 x_5 + a_{54} \theta_4 x_8\\ 
\dot x_6 &= c_1 \left(\xi_6 - x_6\right) u - \theta_2 x_6 + a_{64} \theta_4 x_8\\ 
\dot x_7 &= c_1 \left(\xi_7 - x_7\right) u - \theta_3 x_7 + a_{74} \theta_4 x_8\\
\dot x_8 &= c_1 \left(\xi_8 - x_8\right) u + a_{81} \theta_1 x_5 + a_{82} \theta_2 x_6 + a_{83} \theta_3 x_7 - \theta_4 x_8 \\
\dot x_9 &= c_1 \left(\xi_9 - x_9\right) u \\
\dot x_{10} &= c_{15} x_{10}^3 + c_{16} x_{10}^2 x_{11} + c_{17} x_{10} x_{11}^2 + c_{18} x_{10}^2 + c_{19} x_{10} x_{11} + c_{20} x_{10} + c_5 x_1
\label{R4:x9-with-y1} \\
\dot x_{11} &= c_{17} x_{11}^3 + c_{16} x_{10} x_{11}^2 + c_{15} x_{10}^2 x_{11} + c_{19} x_{11}^2 + c_{18} x_{10} x_{11} + c_{21} x_{11} + c_5 x_2
\label{R4:x10-with-y1}
\end{align}
\end{subequations}
$a_{ij}$ are the absolute values of the entries of the petersen matrix given in Tab.~\ref{tab:app:petersen-ADM1-R4}. $i$ denotes the column (component) and $j$ the row (process). For brevity, only those entries with an absolute value $\ne 1$ or 0 were denoted specifically.

Measurements: 
\begin{subequations}
\label{eq:ADM1-R4-y}
\begin{align}
y_1 &= \dot{V}_g = c_6 x_{10}^2 + c_7 x_{10} x_{11} + c_8 x_{11}^2 + c_9 x_{10} + c_{10} x_{11} + c_{11} \label{R4-V-p-law-controlNotation}\\
y_2 &= p_\mathrm{ch4} = c_{12} x_{10} \\
y_3 &= p_\mathrm{co2} = c_{13} x_{11}  \label{R4:y3} \\
y_4 &= S_\mathrm{IN} = x_3 \\
y_5 &= TS = 1 - \frac{1}{c_{14}} x_4 \\
y_6 &= VS = 1 - \frac{1}{c_{14} - x_4} x_9
\end{align}
\end{subequations}
Tab.~\ref{tab:R4-param-nomenclature} summarizes the aggregated parameters $c_i$ and the time-variant parameters $\theta_i$ used for ADM1-R4 and its submodels. 
\begin{table}[tb]
	\vspace*{-7pt}
	\renewcommand{\arraystretch}{1.1}
	\centering
	\caption{Aggregated and time-variant parameters and notation of ADM1-R4 and its submodels.}
	\label{tab:R4-param-nomenclature}
	\begin{tabular}{cl} 
		\toprule
		Aggregated notation & Notation by Weinrich and Nelles \citep{Weinrich2021b}\\
		\midrule
		$u$ & $\dot V_f$ \\
		\midrule
		$\theta_1$ & $k_\mathrm{ch}$ \\
		$\theta_2$ & $k_\mathrm{pr}$ \\
		$\theta_3$ & $k_\mathrm{li}$ \\
		$\theta_4$ & $k_\mathrm{dec}$ \\
		\midrule
		$c_1$ & $V_l^{-1}$ \\ 
		$c_2$ & $k_L a$\\
		$c_3$ & $k_L a \, K_{H,\mathrm{ch4}} \bar{R} T$ \\
		$c_4$  & $k_L a \, K_{H,\mathrm{co2}} \bar{R} T$ \\
		$c_5$ & $k_L a \, V_l V_g^{-1} $ \\
		$c_6$ & $k_p p_0^{-1} \left(\bar R T \bar{M}_\mathrm{ch4}^{-1}\right)^2$ \\
		$c_7$ & $2 k_p p_0^{-1} \left(\bar R T\right)^2 \bar{M}_\mathrm{ch4}^{-1} \bar{M}_\mathrm{co2}^{-1}$ \\
		$c_8$ & $k_p p_0^{-1} \left(\bar R T \bar{M}_\mathrm{co2}^{-1}\right)^2$ \\
		$c_9$  & $k_p p_0^{-1} \bar R T \bar{M}_\mathrm{ch4}^{-1} \left(2 p_\mathrm{h2o} - p_0\right)$\\		
		$c_{10}$ & $k_p p_0^{-1} \bar R T \bar{M}_\mathrm{co2}^{-1} \left(2 p_\mathrm{h2o} - p_0\right)$ \\
		$c_{11}$ & $k_p p_0^{-1} \left(p_\mathrm{h2o} - p_0\right) p_\mathrm{h2o}$ \\
		$c_{12}$  & $\bar R T \bar{M}_\mathrm{ch4}^{-1}$ \\
		$c_{13}$ & $\bar R T \bar{M}_\mathrm{co2}^{-1}$ \\	
		$c_{14}$ & $\rho_l$ \\
		$c_{15}$ & $- k_p p_0^{-1} V_g^{-1} \left(\bar R T \bar{M}_\mathrm{ch4}^{-1}\right)^2$ \\
		$c_{16}$ & $-2 k_p p_0^{-1}V_g^{-1} \left(\bar R T\right)^2 \bar{M}_\mathrm{ch4}^{-1} \bar{M}_\mathrm{co2}^{-1}$\\
		$c_{17}$ & $-k_p p_0^{-1}V_g^{-1} \left(\bar R T \bar{M}_\mathrm{co2}^{-1}\right)^2$\\
		$c_{18}$ & $-k_p p_0^{-1}V_g^{-1} \bar R T \bar{M}_\mathrm{ch4}^{-1} \left(2 p_\mathrm{h2o} - p_0\right)$\\
		$c_{19}$ & $-k_p p_0^{-1}V_g^{-1} \bar R T \bar{M}_\mathrm{co2}^{-1} \left(2 p_\mathrm{h2o} - p_0\right)$\\
		$c_{20}$ & $-k_L a \, V_l V_g^{-1} K_{H,\mathrm{ch4}} \bar{R} T - k_p p_0^{-1}V_g^{-1} \left(p_\mathrm{h2o} - p_0\right) p_\mathrm{h2o}$\\
		$c_{21}$ & $-k_L a \, V_l V_g^{-1} K_{H,\mathrm{co2}} \bar{R} T - k_p p_0^{-1}V_g^{-1} \left(p_\mathrm{h2o} - p_0\right) p_\mathrm{h2o}$\\
		$c_{22}$ & $V_l V_g^{-1}$\\
		$c_{23}$ & $-k_p p_0^{-1}V_g^{-1} \left(p_\mathrm{h2o} - p_0\right) p_\mathrm{h2o}$\\
		\bottomrule
	\end{tabular}
\end{table}
\subsubsection*{BMR4+B}
The state vector remains as in \eqref{eq:R4:stateVector}. Differential equations: 
\begin{subequations} \label{BMR4+B-controlNotation}
	\begin{align}
	\dot x_1 &= c_1 \left(\xi_1 - x_1\right) u + a_{11} \theta_1 x_5 + a_{12} \theta_2 x_6 + a_{13} \theta_3 x_7 + a_{14} \theta_4 x_8 - c_2 x_1 + c_3 x_{10} \\
	\dot x_2 &= c_1 \left(\xi_2 - x_2\right) u + a_{21} \theta_1 x_5 + a_{22} \theta_2 x_6 + a_{23} \theta_3 x_7 + a_{24} \theta_4 x_8 - c_2 x_2 + c_4 x_{11} \\
	\dot x_3 &= c_1 \left(\xi_3 - x_3\right) u - a_{31} \theta_1 x_5 + a_{32} \theta_2 x_6 - a_{33} \theta_3 x_7 + a_{34} \theta_4 x_8 \\
	\dot x_4 &= c_1 \left(\xi_4 - x_4\right) u - a_{41} \theta_1 x_5 - a_{42} \theta_2 x_6 - a_{43} \theta_3 x_7 - a_{44} \theta_4 x_8\\ 
	\dot x_5 &= c_1 \left(\xi_5 - x_5\right) u - \theta_1 x_5 \\ 
	\dot x_6 &= c_1 \left(\xi_6 - x_6\right) u - \theta_2 x_6 \\ 
	\dot x_7 &= c_1 \left(\xi_7 - x_7\right) u - \theta_3 x_7 \\
	\dot x_8 &= c_1 \left(\xi_8 - x_8\right) u + a_{81} \theta_1 x_5 + a_{82} \theta_2 x_6 + a_{83} \theta_3 x_7 - a_{84} \theta_4 x_8 \\
	\dot x_9 &= c_1 \left(\xi_9 - x_9\right) u \\
	\dot x_{10} &= c_{15} x_{10}^3 + c_{16} x_{10}^2 x_{11} + c_{17} x_{10} x_{11}^2 + c_{18} x_{10}^2 + c_{19} x_{10} x_{11} + c_{20} x_{10} + c_5 x_1 \\
	\dot x_{11} &= c_{17} x_{11}^3 + c_{16} x_{10} x_{11}^2 + c_{15} x_{10}^2 x_{11} + c_{19} x_{11}^2 + c_{18} x_{10} x_{11} + c_{21} x_{11} + c_5 x_2.
	\end{align}
\end{subequations}
Note that omitting model part A entails modified stoichiometric constants $a_{ij}$ as given in Tab.~\ref{tab:app:petersen-BMR4+B}. The other parameters remain the same and are summarized in Tab.\ref{tab:R4-param-nomenclature}. The measurements remain as in \eqref{eq:ADM1-R4-y}. 
\subsubsection*{BMR4+A}
State vector: 
\begin{equation}
\label{eq:BMR4+A:x}
x = \left[S_\mathrm{IN}, S_\mathrm{h2o}, X_\mathrm{ch}, X_\mathrm{pr}, X_\mathrm{li}, X_\mathrm{bac}, X_\mathrm{ash}, S_\mathrm{ch4,gas}, S_\mathrm{co2,gas}\right]^T
\end{equation}
Differential equations, where $a_{ij}$ are given in Tab.~\ref{tab:app:petersen-ADM1-R4} and the remaining parameters in Tab.\ref{tab:R4-param-nomenclature}: 
\begin{subequations} \label{BMR4+A-controlNotation}
	\begin{align}
	\dot x_1 &= c_1 \left(\xi_1 - x_1\right) u - a_{31} \theta_1 x_3 + a_{32} \theta_2 x_4 - a_{33} \theta_3 x_5 \\
	\dot x_2 &= c_1 \left(\xi_2 - x_2\right) u - a_{41} \theta_1 x_3 - a_{42} \theta_2 x_4 - a_{43} \theta_3 x_5 \\ 
	\dot x_3 &= c_1 \left(\xi_3 - x_3\right) u - \theta_1 x_3 + a_{54} \theta_4 x_6\\ 
	\dot x_4 &= c_1 \left(\xi_4 - x_4\right) u - \theta_2 x_4 + a_{64} \theta_4 x_6\\ 
	\dot x_5 &= c_1 \left(\xi_5 - x_5\right) u - \theta_3 x_5 + a_{74} \theta_4 x_6\\
	\dot x_6 &= c_1 \left(\xi_6 - x_6\right) u + a_{81} \theta_1 x_3 + a_{82} \theta_2 x_4 + a_{83} \theta_3 x_5 - \theta_4 x_6 \\
	\dot x_7 &= c_1 \left(\xi_7 - x_7\right) u \\
	\dot x_8 &= c_{22} a_{11} \theta_1 x_3 + c_{22} a_{12} \theta_2 x_4 + c_{22} a_{13} \theta_3 x_5 + c_{15} x_8^3 + c_{16} x_8^2 x_9 + c_{17} x_8 x_9^2 + c_{18} x_8^2 + \notag \\
	&\text{\quad \,} + c_{19} x_8 x_9 + c_{23} x_8 \\
	\dot x_9 &= c_{22} a_{21} \theta_1 x_3 + c_{22} a_{22} \theta_2 x_4 + c_{22} a_{23} \theta_3 x_5 + c_{17} x_9^3 + c_{16} x_8 x_9^2 + c_{15} x_8^2 x_9 + c_{19} x_9^2 + \notag \\
	&\text{\quad \,} + c_{18} x_8 x_9 + c_{23} x_9
	\end{align}
\end{subequations}
Measurements: 
\begin{subequations}
	\label{eq:BMR4+A-y}
	\begin{align}
	y_1 &= \dot V_g = c_6 x_8^2 + c_7 x_8 x_9 + c_8 x_9^2 + c_9 x_8 + c_{10} x_9 + c_{11}, \\
	y_2 &= p_\mathrm{ch4} = c_{12} x_8, \\
	y_3 &= p_\mathrm{co2} = c_{13} x_9, \\
	y_4 &= S_\mathrm{IN} = x_1, \\
	y_5 &= TS = 1 - \frac{1}{c_{14}} x_2, \\
	y_6 &= VS = 1 - \frac{1}{c_{14} - x_2} x_7.
	\end{align}
\end{subequations}
\subsubsection*{BMR4}
The state vector remains as in \eqref{eq:BMR4+A:x}. Differential equations: 
\begin{subequations} \label{BMR4-controlNotation}
	\begin{align}
	\dot x_1 &= c_1 \left(\xi_1 - x_1\right) u - a_{31} \theta_1 x_3 + a_{32} \theta_2 x_4 - a_{33} \theta_3 x_5 + a_{34} \theta_4 x_6 \\
	\dot x_2 &= c_1 \left(\xi_2 - x_2\right) u - a_{41} \theta_1 x_3 - a_{42} \theta_2 x_4 - a_{43} \theta_3 x_5 - a_{44} \theta_4 x_6\\ 
	\dot x_3 &= c_1 \left(\xi_3 - x_3\right) u - \theta_1 x_3\\ 
	\dot x_4 &= c_1 \left(\xi_4 - x_4\right) u - \theta_2 x_4\\ 
	\dot x_5 &= c_1 \left(\xi_5 - x_5\right) u - \theta_3 x_5\\
	\dot x_6 &= c_1 \left(\xi_6 - x_6\right) u + a_{81} \theta_1 x_3 + a_{82} \theta_2 x_4 + a_{83} \theta_3 x_5 - a_{84} \theta_4 x_6 \\
	\dot x_7 &= c_1 \left(\xi_7 - x_7\right) u \\
	\dot x_8 &= c_{22} a_{11} \theta_1 x_3 + c_{22} a_{12} \theta_2 x_4 + c_{22} a_{13} \theta_3 x_5 + c_{22} a_{14} \theta_4 x_6 + c_{15} x_8^3 + c_{16} x_8^2 x_9 + \notag \\
	&\text{\quad \,} + c_{17} x_8 x_9^2 + c_{18} x_8^2 + c_{19} x_8 x_9 + c_{23} x_8 \\
	\dot x_9 &= c_{22} a_{21} \theta_1 x_3 + c_{22} a_{22} \theta_2 x_4 + c_{22} a_{23} \theta_3 x_5 + c_{22} a_{24} \theta_4 x_6 + c_{17} x_9^3 + c_{16} x_8 x_9^2 + \notag \\
	&\text{\quad \,} + c_{15} x_8^2 x_9 + c_{19} x_9^2 + c_{18} x_8 x_9 + c_{23} x_9
	\end{align}
\end{subequations}
$a_{ij}$ are given in Tab.~\ref{tab:app:petersen-BMR4+B} and the remaining parameters in Tab.\ref{tab:R4-param-nomenclature}. The measurements remain as in \eqref{eq:BMR4+A-y}. 
\begin{landscape}
	\scriptsize
	\renewcommand{\arraystretch}{2.1}
	\setlength{\tabcolsep}{1pt}
	\centering
	\begin{longtable}{ll*{11}{C{1.2cm}}l} 
		\caption{Petersen matrix of ADM1-R4, derived from \citep{SorenWeinrich2017}.} \label{tab:app:petersen-ADM1-R4} \\
		\toprule \\ [-12mm]
		\multicolumn{2}{l}{\textbf{Component i $\rightarrow$}}  & 1  & 2 & 3 & 4 & 5 & 6 & 7 & 8 & 9 & 10 & 11 \\[-4mm]
		\textbf{j} & \textbf{Process $\downarrow$}&   $S_\mathrm{ch4}$ & $S_\mathrm{IC}$ & $S_\mathrm{IN}$ & $S_\mathrm{h2o}$ & $X_\mathrm{ch}$ & $X_\mathrm{pr}$ & $X_\mathrm{li}$ & $X_\mathrm{bac}$ & $X_\mathrm{ash}$ & $S_\mathrm{ch4,gas}$ & $S_\mathrm{co2,gas}$ &  \textbf{Process rate} $r_j$\\ [-0.3mm] 
		\midrule \\ [-11mm]
		1 & Fermentation $X_\mathrm{ch}$ & 0.2482 & 0.6809 & -0.0207 & -0.0456 & -1 & & & 0.1372 & & & & $\theta_1 \,  x_5$\\ [-4mm]
		2 & Fermentation $X_\mathrm{pr}$ & 0.3221 & 0.7954 & 0.1689 & -0.4588 & & -1 & & 0.1723 & & & & $\theta_2 \,  x_6$\\[-4mm] 
		3 & Fermentation $X_\mathrm{li}$ & 0.6393 & 0.5817 & -0.0344 & -0.4152 & & & -1 & 0.2286 & & & & $\theta_3 \,  x_7$\\ [-1mm]
		\hdashline[0.6pt/1.8pt] \\ [-11mm]
		4 & Decay $X_\mathrm{bac}$ & & & & & 0.18 & 0.77 & 0.05 & -1 & & & & $\theta_4 \,  x_8$\\[-1mm]
		\hdashline[0.6pt/1.8pt] \\ [-11mm]
		5 & Phase transition $S_\mathrm{ch4}$ & -1 & & & & & & & & & $c_{22}$ & & $c_2 x_1 - c_3 x_{10}$\\[-4mm]
		6 & Phase transition $S_\mathrm{IC}$ & & -1 & & & & & & & & & $c_{22}$ & $c_2 x_2 - c_4 x_{11}$\\ [-1mm]
		\bottomrule
	\end{longtable}
	\begin{longtable}{ll*{11}{C{1.2cm}}l} 
		\caption{Petersen matrix of BMR4+B, derived from \citep{SorenWeinrich2017}.} \label{tab:app:petersen-BMR4+B} \\
		\toprule \\ [-12mm]
		\multicolumn{2}{l}{\textbf{Component i $\rightarrow$}}  & 1 & 2 & 3 & 4 & 5 & 6 & 7 & 8 & 9 & 10 & 11 \\[-4mm]
		\textbf{j} & \textbf{Process $\downarrow$}& $S_\mathrm{ch4}$ & $S_\mathrm{IC}$ & $S_\mathrm{IN}$ & $S_\mathrm{h2o}$ & $X_\mathrm{ch}$ & $X_\mathrm{pr}$ & $X_\mathrm{li}$ & $X_\mathrm{bac}$ & $X_\mathrm{ash}$ & $S_\mathrm{ch4,gas}$ & $S_\mathrm{co2,gas}$ &  \textbf{Process rate} $r_j$\\ [-0.3mm] 
		\midrule \\ [-11mm]
		1 & Fermentation $X_\mathrm{ch}$ & 0.2482 & 0.6809 & -0.0207 & -0.0456 & -1 & & & 0.1372 & & & & $\theta_1 \,  x_5$\\ [-4mm]
		2 & Fermentation $X_\mathrm{pr}$ & 0.3221 & 0.7954 &  0.1689 & -0.4588 & & -1 & & 0.1723 & & & & $\theta_2 \,  x_6$\\[-4mm] 
		3 & Fermentation $X_\mathrm{li}$ & 0.6393 & 0.5817 & -0.0344 & -0.4152 & & & -1 & 0.2286 & & & & $\theta_3 \,  x_7$\\ [-1mm]
		\hdashline[0.6pt/1.8pt] \\ [-11mm]
		4 & Decay $X_\mathrm{bac}$ & 0.3246 & 0.7641 & 0.1246 & -0.3822 & & & & -0.8312 & & & & $\theta_4 \,  x_8$\\[-1mm]
		\hdashline[0.6pt/1.8pt] \\ [-11mm]
		5 & Phase transition $S_\mathrm{ch4}$& -1&  & & & & & &&&$c_{22}$&& $c_2 x_1 - c_3 x_{10}$\\[-4mm]
		6 & Phase transition $S_\mathrm{IC}$& & -1 & & & & & &&&&$c_{22}$& $c_2 x_2 - c_4 x_{11}$\\ [-1mm]
		\bottomrule
	\end{longtable}
\end{landscape}

\subsection*{ADM1-R3 Models}
Aside from ADM1-R3, the following models were analyzed: BMR3+AC, BMR3+BC, and BMR3+ABC. 
\subsubsection*{ADM1-R3}
State vector: 
\begin{align}
x = \big[&S_\mathrm{ac}, S_\mathrm{ch4}, S_\mathrm{IC}, S_\mathrm{IN}, S_\mathrm{h2o}, X_\mathrm{ch}, X_\mathrm{pr}, X_\mathrm{li}, X_\mathrm{bac}, X_\mathrm{ac}, X_\mathrm{ash},... \notag \\
&S_\mathrm{ion}, S_\mathrm{ac^-}, S_\mathrm{hco3^-}, S_\mathrm{nh3}, S_\mathrm{ch4,gas}, S_\mathrm{co2,gas}\big]^T \label{eq:R3:stateVector}
\end{align}
Differential equations: 
\begin{subequations} \label{eq:R3:x}
	\small
	\begin{align}
	\dot x_1 &= c_1 \left(\xi_1 - x_1\right) u + a_{11} \theta_1 x_6 + a_{12} \theta_2 x_7 + a_{13} \theta_3 x_8 - a_{14} \theta_5 \frac{x_1 \, x_{10}}{\theta_6 + x_1} I_\mathrm{ac}  \\
	\dot x_2 &= c_1 \left(\xi_2 - x_2\right) u + a_{21} \theta_1 x_6 + a_{22} \theta_2 x_7 + a_{23} \theta_3 x_8 - c_5 x_2 + c_6 x_{16} + a_{24} \theta_5  \frac{x_1 \, x_{10}}{\theta_6 + x_1} I_\mathrm{ac} \label{eq:R3:x:Sch4} \\
	\dot x_3 &= c_1 \left(\xi_3 - x_3\right) u + a_{31} \theta_1 x_6 + a_{32} \theta_2 x_7 - a_{33} \theta_3 x_8 - c_5 x_3 + c_5 x_{14} + c_7 x_{17} + a_{34} \theta_5 \frac{x_1 \, x_{10}}{\theta_6 + x_1} I_\mathrm{ac} \label{eq:R3:x:Sco2} \\
	\dot x_4 &= c_1 \left(\xi_4 - x_4\right) u - a_{41} \theta_1 x_6 + a_{42} \theta_2 x_7 - a_{43} \theta_3 x_8 - a_{44} \theta_5 \frac{x_1 \, x_{10}}{\theta_6 + x_1} I_\mathrm{ac} \\
	\dot x_5 &= c_1 \left(\xi_5 - x_5\right) u - a_{51} \theta_1 x_6 - a_{52} \theta_2 x_7 - a_{53} \theta_3 x_8 + a_{54} \theta_5 \frac{x_1 \, x_{10}}{\theta_6 + x_1} I_\mathrm{ac} \\
	\dot x_6 &= c_1 \left(\xi_6 - x_6\right) u - \theta_1 x_6 + a_{65} \theta_4 x_9 + a_{66} \theta_4 x_{10} \\
	\dot x_7 &= c_1 \left(\xi_7 - x_7\right) u - \theta_2 x_7 + a_{75} \theta_4 x_9 + a_{76} \theta_4 x_{10} \\ 
	\dot x_8 &= c_1 \left(\xi_8 - x_8\right) u - \theta_3 x_8 + a_{85} \theta_4 x_9 + a_{86} \theta_4 x_{10} \\ 
	\dot x_9 &= c_1 \left(\xi_9 - x_9\right) u + a_{91} \theta_1 x_6 + a_{92} \theta_2 x_7 + a_{93} \theta_3 x_8 - \theta_4 x_9 \\
	\dot x_{10} &= c_1 \left(\xi_{10} - x_{10}\right) u - \theta_4 x_{10} + \theta_5 \frac{x_1 \, x_{10}}{\theta_6 + x_1} I_\mathrm{ac} \\
	\dot x_{11} &= c_1 \left(\xi_{11} - x_{11}\right) u \\
	\dot x_{12} &= c_1 \left(\xi_{12} - x_{12}\right) u \\
	\dot x_{13} &= c_{29} \left(x_1 - x_{13}\right) - c_9 x_{13} S_\mathrm{H^+} \\
	\dot x_{14} &= c_{30} \left(x_3 - x_{14}\right) - c_{10} x_{14} S_\mathrm{H^+} \\
	\dot x_{15} &= c_{31} \left(x_4 - x_{15}\right) - c_{11} x_{15} S_\mathrm{H^+} \\
	\dot x_{16} &= c_{22} x_{16}^3 + c_{23} x_{16}^2 x_{17} + c_{24} x_{16} x_{17}^2 + c_{25} x_{16}^2 + c_{26} x_{16} x_{17} + c_{12} x_2 + c_{27} x_{16} \label{R3:x16-with-y1}  \\
	\dot x_{17} &= c_{24} x_{17}^3 + c_{23} x_{16} x_{17}^2 + c_{22} x_{16}^2 x_{17} + c_{26} x_{17}^2 + c_{25} x_{16} x_{17} + c_{12} x_3 - c_{12} x_{14} + c_{28} x_{17} \label{R3:x17-with-y1}
	\end{align}
\end{subequations}
$I_\mathrm{ac}$ and $S_\mathrm{H^+}$ are defined as  
\begin{align}
I_\mathrm{ac} &= \frac{c_3}{c_3 + S_\mathrm{H^+}^{c_2}} \frac{x_4}{x_4 + c_8} \frac{\theta_7}{\theta_7 + x_{15}} \label{eq:R3:Iac} \\
S_\mathrm{H^+} &= -\frac{\Phi}{2} + \frac{1}{2} \sqrt{\Phi^2 + c_4}  \text{ , where } \label{R3-SH+-x}\\
\Phi &= x_{12} + \frac{x_4 - x_{15}}{17} - \frac{x_{14}}{44} - \frac{x_{13}}{60}.\label{eq:R3:Phi}
\end{align}
The absolute values of the stoichiometric coefficients $a_{ij}$ are given in the petersen matrix of ADM1-R3, Tab.~\ref{tab:app:petersen-ADM1-R3}. Note that for all ADM1-R3 models, definition and indexing of parameters and stoichiometric constants is independent from all ADM1-R4 models.

Measurements: 
\begin{subequations}
	\label{eq:ADM1-R3-y}
	\begin{align}
	y_1 &= \dot{V}_g = c_{13} x_{16}^2 + c_{14} x_{16} x_{17} + c_{15} x_{17}^2 + c_{16} x_{16} + c_{17} x_{17} + c_{18} \\
	y_2 &= p_\mathrm{ch4} = c_{19} x_{16} \label{R3-pch4} \\
	y_3 &= p_\mathrm{co2} = c_{20} x_{17} \label{R3-pco2} \\
	y_4 &= pH = -\log_{10} S_\mathrm{H^+} \label{eq:R3:y:pH}\\
	y_5 &= S_\mathrm{IN} = x_4 \\
	y_6 &= TS = 1 - \frac{1}{c_{21}} x_5, \label{R3-TS} \\
	y_7 &= VS = 1 - \frac{1}{c_{21} - x_5} x_{11} \label{R3-VS} \\
	y_8 &= S_\mathrm{ac} = x_1
	\end{align}
\end{subequations}
Tab.~\ref{tab:R3-param-nomenclature} summarizes the aggregated parameters $c_i$ and the time-variant parameters $\theta_i$ used for ADM1-R3 and all of its submodels. 
\begin{table}[htb]
	\vspace*{-7pt}
	\renewcommand{\arraystretch}{1.1}
	\centering
	\caption{Aggregated and time-variant parameters and notation of ADM1-R3 and its submodels.}
	\label{tab:R3-param-nomenclature}
	\begin{tabular}{cl} 
		\toprule
		Aggregated notation & Notation by Weinrich and Nelles \citep{Weinrich2021b}\\
		\midrule
		$u$ & $\dot V_f$ \\
		\midrule
		$\theta_1$ & $k_\mathrm{ch}$ \\
		$\theta_2$ & $k_\mathrm{pr}$ \\
		$\theta_3$ & $k_\mathrm{li}$ \\
		$\theta_4$ & $k_\mathrm{dec}$ \\
		$\theta_5$ & $\mu_{m,\mathrm{ac}}$ \\
		$\theta_6$ & $K_{S,\mathrm{ac}}$ \\
		$\theta_7$ & $K_{I,\mathrm{nh3}}$\\
		\midrule
		$c_1$ & $V_l^{-1}$ \\ 
		$c_2$ & $n_\mathrm{ac}$ \\
		$c_3$ & $10^{- \frac{3}{2} \frac{pH_{UL,\mathrm{ac}} + pH_{LL,\mathrm{ac}}}{pH_{UL,\mathrm{ac}} - pH_{LL,\mathrm{ac}}}}$\\
		$c_4$ & $4 K_W$ \\
		$c_5$ & $k_L a$\\
		$c_6$ & $k_L a \, K_{H,\mathrm{ch4}} \bar{R} T$ \\ 
		$c_7$ & $k_L a \, K_{H,\mathrm{co2}} \bar{R} T$ \\	
		$c_8$ & $K_{S,\mathrm{IN}}$ \\  
		$c_9$ & $k_{AB,\mathrm{ac}}$ \\ 		
		$c_{10}$ & $k_{AB,\mathrm{co2}}$ \\
		$c_{11}$ & $k_{AB,\mathrm{IN}}$ \\		
		%
		$c_{12}$ & $k_L a \, V_l V_g^{-1}$ \\
		$c_{13}$ & $k_p p_0^{-1} \left(\bar R T \bar{M}_\mathrm{ch4}^{-1}\right)^2$ \\
		$c_{14}$ & $2 k_p p_0^{-1} \left(\bar R T\right)^2 \bar{M}_\mathrm{ch4}^{-1} \bar{M}_\mathrm{co2}^{-1}$ \\	
		$c_{15}$ & $k_p p_0^{-1} \left(\bar R T \bar{M}_\mathrm{co2}^{-1}\right)^2$ \\
		$c_{16}$ & $k_p p_0^{-1} \bar R T \bar{M}_\mathrm{ch4}^{-1} \left(2 p_\mathrm{h2o} - p_0\right)$ \\
		$c_{17}$ & $k_p p_0^{-1} \bar R T \bar{M}_\mathrm{co2}^{-1} \left(2 p_\mathrm{h2o} - p_0\right)$ \\
		$c_{18}$ & $k_p p_0^{-1} \left(p_\mathrm{h2o} - p_0\right) p_\mathrm{h2o}$ \\
		$c_{19}$ & $\bar R T \bar{M}_\mathrm{ch4}^{-1}$ \\
		$c_{20}$ & $\bar R T \bar{M}_\mathrm{co2}^{-1}$ \\
		$c_{21}$ & $\rho_l$ \\		
		$c_{22}$ & $-k_p V_g^{-1} p_0^{-1} \left(\bar R T \bar{M}_\mathrm{ch4}^{-1}\right)^2 $ \\
		$c_{23}$ & $-2 k_p V_g^{-1} p_0^{-1} \left(\bar R T\right)^2 \bar{M}_\mathrm{ch4}^{-1} \bar{M}_\mathrm{co2}^{-1}$ \\  
		$c_{24}$ & $-k_p V_g^{-1} p_0^{-1} \left(\bar R T \bar{M}_\mathrm{co2}^{-1}\right)^2$ \\ 
		$c_{25}$ & $-k_p V_g^{-1} p_0^{-1} \bar R T \bar{M}_\mathrm{ch4}^{-1} \left(2 p_\mathrm{h2o} - p_0\right)$ \\ 		
		$c_{26}$ & $-k_p V_g^{-1} p_0^{-1} \bar R T \bar{M}_\mathrm{co2}^{-1} \left(2 p_\mathrm{h2o} - p_0\right)$ \\ 
		$c_{27}$ & $- k_L a \, V_l V_g^{-1} K_{H,\mathrm{ch4}} \bar{R} T - k_p V_g^{-1} p_0^{-1} \left(p_\mathrm{h2o} - p_0\right) p_\mathrm{h2o}$ \\ 
		$c_{28}$ & $- k_L a \, V_l V_g^{-1} K_{H,\mathrm{co2}} \bar{R} T - k_p V_g^{-1} p_0^{-1} \left(p_\mathrm{h2o} - p_0\right) p_\mathrm{h2o}$ \\ 
		$c_{29}$ & $k_{AB,\mathrm{ac}} K_{a,\mathrm{ac}}$ \\ 
		$c_{30}$ & $k_{AB,\mathrm{co2}} K_{a,\mathrm{co2}}$ \\  
		$c_{31}$ & $k_{AB,\mathrm{IN}} K_{a,\mathrm{IN}}$ \\ 
		$c_{32}$ & $V_l V_g^{-1}$ \\ 
		$c_{33}$ & $-k_p V_g^{-1} p_0^{-1} \left(p_\mathrm{h2o} - p_0\right) p_\mathrm{h2o}$ \\ 
		\bottomrule
	\end{tabular}
\end{table}
\subsubsection*{BMR3+ABC}
State vector: 
\begin{equation}
\label{eq:BMR3+ABC:x}
x= \left[S_\mathrm{ac}, S_\mathrm{ch4}, S_\mathrm{IC}, S_\mathrm{IN}, S_\mathrm{h2o}, X_\mathrm{ch}, X_\mathrm{pr}, X_\mathrm{li}, X_\mathrm{bac}, X_\mathrm{ac}, X_\mathrm{ash}, S_\mathrm{ch4,gas}, S_\mathrm{co2,gas}\right]^T
\end{equation}
Differential equations:
\begin{subequations} \label{BMR3+ABC-controlNotation}
	\small 
	\begin{align}
	\dot x_1 &= c_1 \left(\xi_1 - x_1\right) u + a_{11} \theta_1 x_6 + a_{12} \theta_2 x_7 + a_{13} \theta_3 x_8 - a_{14} \theta_5 \frac{x_1 \, x_{10}}{\theta_6 + x_1} \, \frac{x_4}{x_4 + c_8} \\
	\dot x_2 &= c_1 \left(\xi_2 - x_2\right) u + a_{21} \theta_1 x_6 + a_{22} \theta_2 x_7 + a_{23} \theta_3 x_8 - c_5 x_2 + c_6 x_{12} + a_{24} \theta_5  \frac{x_1 \, x_{10}}{\theta_6 + x_1} \, \frac{x_4}{x_4 + c_8} \\
	\dot x_3 &= c_1 \left(\xi_3 - x_3\right) u + a_{31} \theta_1 x_6 + a_{32} \theta_2 x_7 - a_{33} \theta_3 x_8 - c_5 x_3 + c_7 x_{13} + a_{34} \theta_5 \frac{x_1 \, x_{10}}{\theta_6 + x_1} \, \frac{x_4}{x_4 + c_8} \\
	\dot x_4 &= c_1 \left(\xi_4 - x_4\right) u - a_{41} \theta_1 x_6 + a_{42} \theta_2 x_7 - a_{43} \theta_3 x_8 - a_{44} \theta_5 \frac{x_1 \, x_{10}}{\theta_6 + x_1} \, \frac{x_4}{x_4 + c_8} \\
	\dot x_5 &= c_1 \left(\xi_5 - x_5\right) u - a_{51} \theta_1 x_6 - a_{52} \theta_2 x_7 - a_{53} \theta_3 x_8 + a_{54} \theta_5 \frac{x_1 \, x_{10}}{\theta_6 + x_1} \, \frac{x_4}{x_4 + c_8} \\
	\dot x_6 &= c_1 \left(\xi_6 - x_6\right) u - \theta_1 x_6 + a_{65} \theta_4 x_9 + a_{66} \theta_4 x_{10} \\
	\dot x_7 &= c_1 \left(\xi_7 - x_7\right) u - \theta_2 x_7 + a_{75} \theta_4 x_9 + a_{76} \theta_4 x_{10} \\ 
	\dot x_8 &= c_1 \left(\xi_8 - x_8\right) u - \theta_3 x_8 + a_{85} \theta_4 x_9 + a_{86} \theta_4 x_{10} \\ 
	\dot x_9 &= c_1 \left(\xi_9 - x_9\right) u + a_{91} \theta_1 x_6 + a_{92} \theta_2 x_7 + a_{93} \theta_3 x_8 - \theta_4 x_9 \\
	\dot x_{10} &= c_1 \left(\xi_{10} - x_{10}\right) u - \theta_4 x_{10} + \theta_5 \frac{x_1 \, x_{10}}{\theta_6 + x_1} \, \frac{x_4}{x_4 + c_8} \\
	\dot x_{11} &= c_1 \left(\xi_{11} - x_{11}\right) u \\
	\dot x_{12} &= c_{22} x_{12}^3 + c_{23} x_{12}^2 x_{13} + c_{24} x_{12} x_{13}^2 + c_{25} x_{12}^2 + c_{26} x_{12} x_{13} + c_{12} x_2 + c_{27} x_{12} \\
	\dot x_{13} &= c_{24} x_{13}^3 + c_{23} x_{12} x_{13}^2 + c_{22} x_{12}^2 x_{13} + c_{26} x_{13}^2 + c_{25} x_{12} x_{13} + c_{12} x_3 + c_{28} x_{13} 
	\end{align}
\end{subequations}
$a_{ij}$ should be taken from Tab.~\ref{tab:app:petersen-ADM1-R3}. The other involved parameters are given in Tab.~\ref{tab:R3-param-nomenclature}.

Measurements:
\begin{subequations}
	\label{eq:BMR3+ABC-y}
	\begin{align}
	y_1 &= \dot V_g = c_{13} x_{12}^2 + c_{14} x_{12} x_{13} + c_{15} x_{13}^2 + c_{16} x_{12} + c_{17} x_{13} + c_{18} \\
	y_2 &= p_\mathrm{ch4} = c_{19} x_{12} \\
	y_3 &= p_\mathrm{co2} = c_{20} x_{13} \\
	y_4 &= S_\mathrm{IN} = x_4 \\
	y_5 &= TS = 1 - \frac{1}{c_{21}} x_5 \\
	y_6 &= VS = 1 - \frac{1}{c_{21} - x_5} x_{11} \\
	y_7 &= S_\mathrm{ac} = x_1
	\end{align}
\end{subequations}
\subsubsection*{BMR3+BC}
The state vector remains as in \eqref{eq:BMR3+ABC:x}. Differential equations: 
\begin{subequations} \label{BMR3+BC-controlNotation}
	\small
	\begin{align}
	\dot x_1 &= c_1 \left(\xi_1 - x_1\right) u + a_{11} \theta_1 x_6 + a_{12} \theta_2 x_7 + a_{13} \theta_3 x_8 - a_{14} \theta_5 \frac{x_1 \, x_{10}}{\theta_6 + x_1} \, \frac{x_4}{x_4 + c_8} +  a_{15} \theta_4 x_9 + \notag \\
	&\text{\quad \,} + a_{16} \theta_4 x_{10}\\
	\dot x_2 &= c_1 \left(\xi_2 - x_2\right) u + a_{21} \theta_1 x_6 + a_{22} \theta_2 x_7 + a_{23} \theta_3 x_8 - c_5 x_2 + c_6 x_{12} + a_{24} \theta_5  \frac{x_1 \, x_{10}}{\theta_6 + x_1} \, \frac{x_4}{x_4 + c_8} + \notag \\
	&\text{\quad \,} + a_{25} \theta_4 x_9 + a_{26} \theta_4 x_{10} \\
	\dot x_3 &= c_1 \left(\xi_3 - x_3\right) u + a_{31} \theta_1 x_6 + a_{32} \theta_2 x_7 - a_{33} \theta_3 x_8 - c_5 x_3 + c_7 x_{13} + a_{34} \theta_5 \frac{x_1 \, x_{10}}{\theta_6 + x_1} \, \frac{x_4}{x_4 + c_8} + \notag \\
	&\text{\quad \,} + a_{35} \theta_4 x_9 + a_{36} \theta_4 x_{10} \\
	\dot x_4 &= c_1 \left(\xi_4 - x_4\right) u - a_{41} \theta_1 x_6 + a_{42} \theta_2 x_7 - a_{43} \theta_3 x_8 - a_{44} \theta_5 \frac{x_1 \, x_{10}}{\theta_6 + x_1} \, \frac{x_4}{x_4 + c_8} + a_{45} \theta_4 x_9 + \notag \\
	&\text{\quad \,} + a_{46} \theta_4 x_{10}\\
	\dot x_5 &= c_1 \left(\xi_5 - x_5\right) u - a_{51} \theta_1 x_6 - a_{52} \theta_2 x_7 - a_{53} \theta_3 x_8 + a_{54} \theta_5 \frac{x_1 \, x_{10}}{\theta_6 + x_1} \, \frac{x_4}{x_4 + c_8} - a_{55} \theta_4 x_9 + \notag \\
	&\text{\quad \,} - a_{56} \theta_4 x_{10}\\
	\dot x_6 &= c_1 \left(\xi_6 - x_6\right) u - \theta_1 x_6 \\
	\dot x_7 &= c_1 \left(\xi_7 - x_7\right) u - \theta_2 x_7 \\ 
	\dot x_8 &= c_1 \left(\xi_8 - x_8\right) u - \theta_3 x_8  \\ 
	\dot x_9 &= c_1 \left(\xi_9 - x_9\right) u + a_{91} \theta_1 x_6 + a_{92} \theta_2 x_7 + a_{93} \theta_3 x_8 - a_{95} \theta_4 x_9 + a_{96} \theta_4 x_{10}\\
	\dot x_{10} &= c_1 \left(\xi_{10} - x_{10}\right) u - \theta_4 x_{10} + \theta_5 \frac{x_1 \, x_{10}}{\theta_6 + x_1} \, \frac{x_4}{x_4 + c_8} \\
	\dot x_{11} &= c_1 \left(\xi_{11} - x_{11}\right) u \\
	\dot x_{12} &= c_{22} x_{12}^3 + c_{23} x_{12}^2 x_{13} + c_{24} x_{12} x_{13}^2 + c_{25} x_{12}^2 + c_{26} x_{12} x_{13} + c_{12} x_2 + c_{27} x_{12} \\
	\dot x_{13} &= c_{24} x_{13}^3 + c_{23} x_{12} x_{13}^2 + c_{22} x_{12}^2 x_{13} + c_{26} x_{13}^2 + c_{25} x_{12} x_{13} + c_{12} x_3 + c_{28} x_{13} 
	\end{align}
\end{subequations}
Note that omitting model part A entails modified stoichiometric constants $a_{ij}$ as given in Tab.~\ref{tab:app:petersen-BMR3+BC}. The other involved parameters remain the same. They are given in Tab.~\ref{tab:R3-param-nomenclature}.

Measurements: 
\begin{subequations}
	\label{eq:BMR3+BC-y}
	\begin{align}
	y_1 &= \dot V_g = c_{13} x_{12}^2 + c_{14} x_{12} x_{13} + c_{15} x_{13}^2 + c_{16} x_{12} + c_{17} x_{13} + c_{18} \\
	y_2 &= p_\mathrm{ch4} = c_{19} x_{12} \\
	y_3 &= p_\mathrm{co2} = c_{20} x_{13} \\
	y_4 &= S_\mathrm{IN} = x_4 \\
	y_5 &= TS = 1 - \frac{1}{c_{21}} x_5 \\
	y_6 &= VS = 1 - \frac{1}{c_{21} - x_5} x_{11} \\
	y_7 &= S_\mathrm{ac} = x_1
	\end{align}
\end{subequations}
\subsubsection*{BMR3+AC}
State vector: 
\begin{equation}
\label{eq:BMR3+AC:x}
x= \left[S_\mathrm{ac}, S_\mathrm{IN}, S_\mathrm{h2o}, X_\mathrm{ch}, X_\mathrm{pr}, X_\mathrm{li}, X_\mathrm{bac}, X_\mathrm{ac}, X_\mathrm{ash}, S_\mathrm{ch4,gas}, S_\mathrm{co2,gas}\right]^T
\end{equation}
Differential equations:
\begin{subequations} \label{BMR3+AC-controlNotation}
	\begin{align}
	\dot x_1 &= c_1 \left(\xi_1 - x_1\right) u + a_{11} \theta_1 x_4 + a_{12} \theta_2 x_5 + a_{13} \theta_3 x_6 - a_{14} \theta_5 \frac{x_1 \, x_8}{\theta_6 + x_1} \, \frac{x_2}{x_2 + c_8} \\
	\dot x_2 &= c_1 \left(\xi_2 - x_2\right) u - a_{41} \theta_1 x_4 + a_{42} \theta_2 x_5 - a_{43} \theta_3 x_6 - a_{44} \theta_5 \frac{x_1 \, x_8}{\theta_6 + x_1} \, \frac{x_2}{x_2 + c_8} \\
	\dot x_3 &= c_1 \left(\xi_3 - x_3\right) u - a_{51} \theta_1 x_4 - a_{52} \theta_2 x_5 - a_{53} \theta_3 x_6 + a_{54} \theta_5 \frac{x_1 \, x_8}{\theta_6 + x_1} \, \frac{x_2}{x_2 + c_8} \\
	\dot x_4 &= c_1 \left(\xi_4 - x_4\right) u - \theta_1 x_4 + a_{65} \theta_4 x_7 + a_{66} \theta_4 x_8 \\
	\dot x_5 &= c_1 \left(\xi_5 - x_5\right) u - \theta_2 x_5 + a_{75} \theta_4 x_7 + a_{76} \theta_4 x_8 \\ 
	\dot x_6 &= c_1 \left(\xi_6 - x_6\right) u - \theta_3 x_6 + a_{85} \theta_4 x_7 + a_{86} \theta_4 x_8 \\ 
	\dot x_7 &= c_1 \left(\xi_7 - x_7\right) u + a_{91} \theta_1 x_4 + a_{92} \theta_2 x_5 + a_{93} \theta_3 x_6 - \theta_4 x_7 \\
	\dot x_8 &= c_1 \left(\xi_8 - x_8\right) u + \theta_5 \frac{x_1 \, x_8}{\theta_6 + x_1} \, \frac{x_2}{x_2 + c_8} - \theta_4 x_8 \\
	\dot x_9 &= c_1 \left(\xi_9 - x_9\right) u \\
	\dot x_{10} &= c_{32} a_{21} \theta_1 x_4 + c_{32} a_{22} \theta_2 x_5 + c_{32} a_{23} \theta_3 x_6 + c_{32} a_{24} \theta_5 \frac{x_1 \, x_8}{\theta_6 + x_1} \, \frac{x_2}{x_2 + c_8} + \notag \\
	&\text{\quad \,} c_{22} x_{10}^3 + c_{23} x_{10}^2 x_{11} + c_{24} x_{10} x_{11}^2 + c_{25} x_{10}^2 + c_{26} x_{10} x_{11} + c_{33} x_{10} \label{eq:BMR3+AC:x:Sch4gas} \\
	\dot x_{11} &= c_{32} a_{31} \theta_1 x_4 + c_{32} a_{32} \theta_2 x_5 - c_{32} a_{33} \theta_3 x_6 + c_{32} a_{34} \theta_5 \frac{x_1 \, x_8}{\theta_6 + x_1} \, \frac{x_2}{x_2 + c_8} + \notag \\
	&\text{\quad \,} c_{24} x_{11}^3 + c_{23} x_{10} x_{11}^2 + c_{22} x_{10}^2 x_{11} + c_{26} x_{11}^2 + c_{25} x_{10} x_{11} + c_{33} x_{11} \label{eq:BMR3+AC:x:Sco2gas}
	\end{align}
\end{subequations}
$a_{ij}$ should be taken from Tab.~\ref{tab:app:petersen-ADM1-R3}. The involved parameters remain the same. They are given in Tab.~\ref{tab:R3-param-nomenclature}. Measurements:
\begin{subequations}
	\label{eq:BMR3+AC-y}
	\begin{align} 
	y_1 &= \dot V_g = c_{13} x_{10}^2 + c_{14} x_{10} x_{11} + c_{15} x_{11}^2 + c_{16} x_{10} + c_{17} x_{11} + c_{18} \\
	y_2 &= p_\mathrm{ch4} = c_{19} x_{10} \\
	y_3 &= p_\mathrm{co2} = c_{20} x_{11} \\
	y_4 &= S_\mathrm{IN} = x_2 \\
	y_5 &= TS = 1 - \frac{1}{c_{21}} x_3 \\
	y_6 &= VS = 1 - \frac{1}{c_{21} - x_3} x_9 \\
	y_7 &= S_\mathrm{ac} = x_1
	\end{align}
\end{subequations}
\subsubsection*{BMR3+A}
The state vector remains as in \eqref{eq:BMR3+AC:x}. Differential equations: 
\begin{subequations} \label{BMR3+A-controlNotation}
	\begin{align}
	\dot x_1 &= c_1 \left(\xi_1 - x_1\right) u + a_{11} \theta_1 x_4 + a_{12} \theta_2 x_5 + a_{13} \theta_3 x_6 - a_{14} \theta_5 \frac{x_1 \, x_8}{\theta_6 + x_1} \\
	\dot x_2 &= c_1 \left(\xi_2 - x_2\right) u - a_{41} \theta_1 x_4 + a_{42} \theta_2 x_5 - a_{43} \theta_3 x_6 - a_{44} \theta_5 \frac{x_1 \, x_8}{\theta_6 + x_1} \\
	\dot x_3 &= c_1 \left(\xi_3 - x_3\right) u - a_{51} \theta_1 x_4 - a_{52} \theta_2 x_5 - a_{53} \theta_3 x_6 + a_{54} \theta_5 \frac{x_1 \, x_8}{\theta_6 + x_1} \\
	\dot x_4 &= c_1 \left(\xi_4 - x_4\right) u - \theta_1 x_4 + a_{65} \theta_4 x_7 + a_{66} \theta_4 x_8 \\
	\dot x_5 &= c_1 \left(\xi_5 - x_5\right) u - \theta_2 x_5 + a_{75} \theta_4 x_7 + a_{76} \theta_4 x_8 \\ 
	\dot x_6 &= c_1 \left(\xi_6 - x_6\right) u - \theta_3 x_6 + a_{85} \theta_4 x_7 + a_{86} \theta_4 x_8 \\ 
	\dot x_7 &= c_1 \left(\xi_7 - x_7\right) u + a_{91} \theta_1 x_4 + a_{92} \theta_2 x_5 + a_{93} \theta_3 x_6 - \theta_4 x_7 \\
	\dot x_8 &= c_1 \left(\xi_8 - x_8\right) u + \theta_5 \frac{x_1 \, x_8}{\theta_6 + x_1} - \theta_4 x_8 \\
	\dot x_9 &= c_1 \left(\xi_9 - x_9\right) u \\
	\dot x_{10} &= c_{32} a_{21} \theta_1 x_4 + c_{32} a_{22} \theta_2 x_5 + c_{32} a_{23} \theta_3 x_6 + c_{32} a_{24} \theta_5 \frac{x_1 \, x_8}{\theta_6 + x_1} + \notag \\
	&\text{\quad \,} c_{22} x_{10}^3 + c_{23} x_{10}^2 x_{11} + c_{24} x_{10} x_{11}^2 + c_{25} x_{10}^2 + c_{26} x_{10} x_{11} + c_{33} x_{10} \\
	\dot x_{11} &= c_{32} a_{31} \theta_1 x_4 + c_{32} a_{32} \theta_2 x_5 - c_{32} a_{33} \theta_3 x_6 + c_{32} a_{34} \theta_5 \frac{x_1 \, x_8}{\theta_6 + x_1} + \notag \\
	&\text{\quad \,} c_{24} x_{11}^3 + c_{23} x_{10} x_{11}^2 + c_{22} x_{10}^2 x_{11} + c_{26} x_{11}^2 + c_{25} x_{10} x_{11} + c_{33} x_{11}
	\end{align}
\end{subequations}
$a_{ij}$ should be taken from Tab.~\ref{tab:app:petersen-ADM1-R3}. The involved parameters remain the same. They are given in Tab.~\ref{tab:R3-param-nomenclature}. Measurements are computed according to~\eqref{eq:BMR3+AC-y}.
\subsubsection*{BMR3}
The state vector remains as in \eqref{eq:BMR3+AC:x}. Differential equations: 
\begin{subequations} \label{BMR3-controlNotation}
	\begin{align}
	\dot x_1 &= c_1 \left(\xi_1 - x_1\right) u + a_{11} \theta_1 x_4 + a_{12} \theta_2 x_5 + a_{13} \theta_3 x_6 - a_{14} \theta_5 \frac{x_1 \, x_8}{\theta_6 + x_1} + a_{15} \theta_4 x_7 + \notag \\
	&\text{\quad \,} + a_{16} \theta_4 x_8 \\
	\dot x_2 &= c_1 \left(\xi_2 - x_2\right) u - a_{41} \theta_1 x_4 + a_{42} \theta_2 x_5 - a_{43} \theta_3 x_6 - a_{44} \theta_5 \frac{x_1 \, x_8}{\theta_6 + x_1} + a_{45} \theta_4 x_7 + \notag \\
	&\text{\quad \,} + a_{46} \theta_4 x_8 \\
	\dot x_3 &= c_1 \left(\xi_3 - x_3\right) u - a_{51} \theta_1 x_4 - a_{52} \theta_2 x_5 - a_{53} \theta_3 x_6 + a_{54} \theta_5 \frac{x_1 \, x_8}{\theta_6 + x_1} - a_{55} \theta_4 x_7 + \notag \\
	&\text{\quad \,} - a_{56} \theta_4 x_8 \\
	\dot x_4 &= c_1 \left(\xi_4 - x_4\right) u - \theta_1 x_4 \\
	\dot x_5 &= c_1 \left(\xi_5 - x_5\right) u - \theta_2 x_5 \\ 
	\dot x_6 &= c_1 \left(\xi_6 - x_6\right) u - \theta_3 x_6 \\ 
	\dot x_7 &= c_1 \left(\xi_7 - x_7\right) u + a_{91} \theta_1 x_4 + a_{92} \theta_2 x_5 + a_{93} \theta_3 x_6 - a_{95} \theta_4 x_7 + a_{96} \theta_4 x_8\\
	\dot x_8 &= c_1 \left(\xi_8 - x_8\right) u + \theta_5 \frac{x_1 \, x_8}{\theta_6 + x_1} - \theta_4 x_8 \\
	\dot x_9 &= c_1 \left(\xi_9 - x_9\right) u \\
	\dot x_{10} &= c_{32} a_{21} \theta_1 x_4 + c_{32} a_{22} \theta_2 x_5 + c_{32} a_{23} \theta_3 x_6 + c_{32} a_{24} \theta_5 \frac{x_1 \, x_8}{\theta_6 + x_1} + c_{32} a_{25} \theta_4 x_7 + \notag \\
	&\text{\quad \,} + c_{32} a_{26} \theta_4 x_8 + c_{22} x_{10}^3 + c_{23} x_{10}^2 x_{11} + c_{24} x_{10} x_{11}^2 + c_{25} x_{10}^2 + c_{26} x_{10} x_{11} + c_{33} x_{10} \\
	\dot x_{11} &= c_{32} a_{31} \theta_1 x_4 + c_{32} a_{32} \theta_2 x_5 - c_{32} a_{33} \theta_3 x_6 + c_{32} a_{34} \theta_5 \frac{x_1 \, x_8}{\theta_6 + x_1} + c_{32} a_{35} \theta_4 x_7 + \notag \\
	&\text{\quad \,} + c_{32} a_{36} \theta_4 x_8 + c_{24} x_{11}^3 + c_{23} x_{10} x_{11}^2 + c_{22} x_{10}^2 x_{11} + c_{26} x_{11}^2 + c_{25} x_{10} x_{11} + c_{33} x_{11}
	\end{align}
\end{subequations}
$a_{ij}$ should be taken from Tab.~\ref{tab:app:petersen-BMR3+BC}. The involved parameters remain the same. They are given in Tab.~\ref{tab:R3-param-nomenclature}. Measurements are computed according to~\eqref{eq:BMR3+AC-y}.
%
%
\begin{landscape}
	\scriptsize
	\renewcommand{\arraystretch}{2.1}
	\setlength{\tabcolsep}{1pt}
	\centering
	\begin{longtable}{ll*{10}{C{1.2cm}}l} 
		\caption{Petersen matrix of ADM1-R3, derived from \citep{SorenWeinrich2017}.} \label{tab:app:petersen-ADM1-R3} \\
		\toprule \\ [-12mm]
		\multicolumn{2}{l}{\textbf{Component i $\rightarrow$}}  & 1  & 2 & 3 & 4 & 5 & 6 & 7 & 8 & 9 & 10 \\[-4mm]
		\textbf{j} & \textbf{Process $\downarrow$} & $S_\mathrm{ac}$ & $S_\mathrm{ch4}$ & $S_\mathrm{IC}$ & $S_\mathrm{IN}$ & $S_\mathrm{h2o}$ & $X_\mathrm{ch}$ & $X_\mathrm{pr}$ & $X_\mathrm{li}$ & $X_\mathrm{bac}$ & $X_\mathrm{ac}$ &  \textbf{Process rate} $r_j$\\ [-1mm] 
		\midrule \\ [-11mm] 
		1 & Fermentation $X_\mathrm{ch}$ & 0.6555 & 0.0818 & 0.2245 & -0.0169 & -0.0574 & -1 & & & 0.1125 & & $\theta_1 \,  x_6$\\ [-4mm]
		2 & Fermentation $X_\mathrm{pr}$ & 0.9947 & 0.0696 & 0.1029 & 0.1746 & -0.4767 & & -1 & & 0.1349 & & $\theta_2 \,  x_7$\\[-4mm] 
		3 & Fermentation $X_\mathrm{li}$ & 1.7651 & 0.1913 & -0.6472 & -0.0244 & -0.4470 & & & -1 & 0.1621 & & $\theta_3 \,  x_8$\\ [-1mm]
		\hdashline[0.6pt/1.8pt] \\ [-10mm]
		4 & Methanogenesis $S_\mathrm{ac}$ & -26.5447 & 6.7367 & 18.4808 & -0.1506 & 0.4778 & & & & & 1 & $\theta_5 \, \frac{x_1}{\theta_6 + x_1} \, x_{10} \, I_{ac}$\\ 
		\hdashline[0.6pt/1.8pt] \\ [-11mm]
		5 & Decay $X_\mathrm{bac}$ & & & & & & 0.18 & 0.77 & 0.05 & -1 & & $\theta_4 \,  x_9$\\[-4mm]
		6 & Decay $X_\mathrm{ac}$ & & & & & & 0.18 & 0.77 & 0.05 & & -1 & $\theta_4 \,  x_{10}$\\[-1mm]
		\midrule \\ [-11mm]
		\midrule \\ [-11mm]  
		& & 2 & 3 & $\ldots$ & 11 & 12 & 13 & 14 & 15 & 16 & 17 & \\[-4mm]
		& & $S_\mathrm{ch4}$ & $S_\mathrm{IC}$ & & $X_\mathrm{ash}$ & $S_\mathrm{ion}$ & $S_\mathrm{ac^-}$ & $S_\mathrm{hco3^-}$ & $S_\mathrm{nh3}$ & $S_\mathrm{ch4,gas}$ & $S_\mathrm{co2,gas}$ & \\ [-1mm] 		%
		\midrule \\ [-11mm] 
		7 & Dissoziation $S_\mathrm{ac}$ & & & & & & -1 & & & & & $c_{29} \, (x_{13} - x_1) + c_9 \, x_{13} \, S_\mathrm{H^+}$\\ [-4mm]
		8 & Dissoziation $S_\mathrm{IC}$ & & & & & & & -1 & & & & $c_{30} \, (x_{14} - x_3) + c_{10} \, x_{14} \, S_\mathrm{H^+}$\\[-4mm] 
		9 & Dissoziation $S_\mathrm{IN}$ & & & & & & & & -1 & & & $c_{31} \, (x_{15} - x_4) + c_{11} \, x_{15} \, S_\mathrm{H^+}$\\ [-1mm]
		\hdashline[0.6pt/1.8pt] \\ [-11mm]
		10 & Phase transition $S_\mathrm{ch4}$& -1 & & & & & & & & $c_{32}$ & & $c_5 \, x_2 - c_6 \, x_{16}$\\[-4mm]
		11 & Phase transition $S_\mathrm{co2}$& & -1 & & & & & & & & $c_{32}$ & $c_5 \, (x_3 - x_{14}) - c_7 \, x_{17}$\\ [-1mm]
		\bottomrule
	\end{longtable}
\end{landscape}
\begin{landscape}
	\scriptsize
	\renewcommand{\arraystretch}{2.1}
	\setlength{\tabcolsep}{1pt}
	\centering
	\begin{longtable}{ll*{13}{C{1.2cm}}l} 
		\caption{Petersen matrix of BMR3+BC, derived from \citep{SorenWeinrich2017}.} \label{tab:app:petersen-BMR3+BC} \\
		\toprule \\ [-12mm]
		\multicolumn{2}{l}{\textbf{Component i $\rightarrow$}}  & 1 & 2 & 3 & 4 & 5 & 6 & 7 & 8 & 9 & 10 & 11 & 12 & 13 \\[-4mm]
		\textbf{j} & \textbf{Process $\downarrow$} & $S_\mathrm{ac}$ & $S_\mathrm{ch4}$ & $S_\mathrm{IC}$ & $S_\mathrm{IN}$ & $S_\mathrm{h2o}$ & $X_\mathrm{ch}$ & $X_\mathrm{pr}$ & $X_\mathrm{li}$ & $X_\mathrm{bac}$ & $X_\mathrm{ac}$ & $X_\mathrm{ash}$ & $S_\mathrm{ch4,gas}$ & $S_\mathrm{co2,gas}$ &  \textbf{Process rate} $r_j$\\ [-1mm] 
		\midrule \\ [-11mm] 
		1 & Fermentation $X_\mathrm{ch}$ & 0.6555 & 0.0818 & 0.2245 & -0.0169 & -0.0574 & -1 & & & 0.1125 & & & & & $\theta_1 \,  x_6$\\ [-4mm]
		2 & Fermentation $X_\mathrm{pr}$ & 0.9947 & 0.0696 & 0.1029 & 0.1746 & -0.4767 & & -1 & & 0.1349 & & & & & $\theta_2 \,  x_7$\\[-4mm] 
		3 & Fermentation $X_\mathrm{li}$ & 1.7651 & 0.1913 & -0.6472 & -0.0244 & -0.4470 & & & -1 & 0.1621 & & & & & $\theta_3 \,  x_8$\\ [-1mm]
		\hdashline[0.6pt/1.8pt] \\ [-10mm]
		4 & Methanogenesis $S_\mathrm{ac}$ & -26.5447 & 6.7367 & 18.4808 & -0.1506 & 0.4778 & & & & & 1 & & & & $\theta_5 \, \frac{x_1}{\theta_6 + x_1} \, x_{10} \, \frac{x_4}{x_4 + c_8}$\\ 
		\hdashline[0.6pt/1.8pt] \\ [-11mm]
		5 & Decay $X_\mathrm{bac}$ & 0.9722 & 0.0779 & 0.0873 & 0.1301 & -0.3997 & & & & -0.8678 & & & & & $\theta_4 \,  x_9$\\[-4mm]
		6 & Decay $X_\mathrm{ac}$ & 0.9722 & 0.0779 & 0.0873 & 0.1301 & -0.3997 & & & & 0.1322 & -1 & & & & $\theta_4 \,  x_{10}$\\[-1mm]
		\hdashline[0.6pt/1.8pt] \\ [-11mm]
		7 & Phase transition $S_\mathrm{ch4}$& & -1 & & & & & & & & & & $c_{32}$ & & $c_5 \, x_2 - c_6 \, x_{12}$\\[-4mm]
		8 & Phase transition $S_\mathrm{co2}$& & & -1 & & & & & & & & & & $c_{32}$ & $c_5 \, x_3 - c_7 \, x_{13}$\\ [-1mm]
		\bottomrule
	\end{longtable} 
\end{landscape}

\subsection*{Derivations for neglecting individual model parts}
This section details the basic ideas behind neglecting individual model parts for both ADM1-R3 and ADM1-R4.
\subsubsection*{Neglecting model part A - stoichiometric degradation of microbial biomass to macro nutrients}
Stoichiometric pathways of the ADM1 models can easily be derived from the rows of the petersen matrix, Tab.~\ref{tab:app:petersen-ADM1-R3}. In the original ADM1-R3/ADM1-R4, biomass (represented by the states $X_\mathrm{bac}$ and $X_\mathrm{ac}$) is formed during degradation of macro nutrients (carbohydrates, proteins and lipids). Biomass in turn is decomposed into macro nutrients (decay of bacteria). This feedback can be removed by modifying the stoichiometry of biomass. Biomass is thereby not decomposed into macro nutrients, but into the stoichiometric degradation products of the macro nutrients directly. This delivers a modified stoichiometry, affecting the differential equations of $S_\mathrm{ac}$, $S_\mathrm{ch4}$, $S_\mathrm{IC}$, $S_\mathrm{IN}$, $S_\mathrm{h2o}$, $X_\mathrm{ch}$, $X_\mathrm{pr}$, $X_\mathrm{li}$, $X_\mathrm{bac}$, and $X_\mathrm{ac}$ (depending on the used model class). 
\subsubsection*{Neglecting model part B - gas solubility of \chfour and \cotwo}
If \chfour and \cotwo are assumed to be insoluble in the liquid phase, their stoichiometric formation therein (see e.g. \eqref{eq:R3:x:Sch4} and \eqref{eq:R3:x:Sco2}) has to be allocated to the gas phase. Gas transfer from liquid to gas phase is originally modeled by means of Henry's law. The according terms can be cut out, as well as the convection terms. Instead, the terms describing stoichioetric formation of \chfour and \cotwo are normalized with the ratio of liquid and gas volume ($V_l/V_g$) and appear in the differential equations of $S_\mathrm{ch4,gas}$ and $S_\mathrm{co2,gas}$, respectively. This can be seen in e.g. \eqref{eq:BMR3+AC:x:Sch4gas} and \eqref{eq:BMR3+AC:x:Sco2gas}. Computation of the gas volume flow remains as in the case with model part B. 
\subsubsection*{Neglecting model part C - inhibition through nitrogen limitation}
The second factor in the inhibition function $I_\mathrm{ac}$ describes inhibition through nitrogen limitation, see \eqref{eq:R3:Iac}. Neglecting model part~C is achieved by removing this factor, which reduces nonlinearities in the model. However, this does not allow to omit any of the state variables. 
\subsubsection*{Neglecting model part D - inhibition through pH and ammonia}
The full inhibition function $I_\mathrm{ac}$ is a major source of nonlinearity in the ADM1-R3. Neglecting model part~D (inhibition through pH and ammonia) allows to cut out the first and last factor of $I_\mathrm{ac}$, see \eqref{eq:R3:Iac}. Consequently, $S_\mathrm{nh3}$ can be omitted in the state vector. 
\subsubsection*{Neglecting model part E - computation of pH}
Measuring the pH 
allows to infer $S_\mathrm{H^+}$ directly because these two variables are linked via the negative common logarithm, \eqref{eq:R3:y:pH}. Measureing the pH hence allows to interpret the variable $S_\mathrm{H^+}$ as a time-variant parameter (without any associated differential equation). The states $S_\mathrm{ion}$, $S_\mathrm{ac^-}$ and $S_\mathrm{hco3^-}$ ($x_{12}$ to $x_{14}$ in \eqref{eq:R3:stateVector} and \eqref{eq:R3:x}) only appear in the computation of the charge balance $\Phi$, \eqref{eq:R3:Phi} which is required to calculate $S_\mathrm{H^+}$. However, as $S_\mathrm{H^+}$ can be directly determined from pH measurements, the states $S_\mathrm{ion}$, $S_\mathrm{ac^-}$ and $S_\mathrm{hco3^-}$ become redundant. Their respective differential equations can be cut out of the system of equations. The resulting model BMR3+ABCD (not shown here) only incorporates dissociation between ammonium and ammonia. Yet, full inhibition through all three factors of $I_\mathrm{ac}$ (pH, nitrogen limitation and ammonia) are considered.  

\interlinepenalty 10000	

\bibliographystyle{unsrt}	
\clearpage
\bibliography{LiteraturAD}							

\end{document}